\documentclass[aps,prl,reprint,longbibliography,superscriptaddress,showpacs]{revtex4-1}
\usepackage{}
\usepackage{amssymb}
\usepackage{amsmath}
\usepackage{dcolumn}
\usepackage{bm}
\usepackage{graphicx}
\usepackage{mathrsfs}
\usepackage[colorlinks,linkcolor=blue,anchorcolor=blue,citecolor=blue,urlcolor=black]%
{hyperref}

\begin{document}
\title{ Hybrid quantum device with  nitrogen-vacancy centers in diamond coupled to  carbon nanotubes}
\author{Peng-Bo  Li}
\email{lipengbo@mail.xjtu.edu.cn}

\affiliation {Center for Emergent Matter Science, RIKEN, Saitama 351-0198, Japan}
\affiliation {Department of Applied Physics, Xi'an Jiaotong University, Xi'an
710049, China}

\author{Ze-Liang Xiang}
\affiliation {Vienna Center for Quantum Science and Technology, Atominstitut, TU Wien, 1020 Vienna, Austria}

\author{Peter Rabl}
\affiliation {Vienna Center for Quantum Science and Technology, Atominstitut, TU Wien, 1020 Vienna, Austria}

\author{Franco  Nori}
\affiliation {Center for Emergent Matter Science, RIKEN, Saitama 351-0198, Japan}
\affiliation{Department of Physics, The University of Michigan, Ann Arbor, Michigan 48109-1040, USA}

\begin{abstract}
We show that NV centers in diamond interfaced
with a suspended carbon nanotube carrying a dc current  can facilitate a  spin-nanomechanical hybrid device.
We demonstrate that strong magnetomechanical interactions between  a single  NV spin and the vibrational mode of the suspended  nanotube  can be engineered and dynamically tuned by external control over the system parameters. This spin-nanomechanical setup with strong, \emph{intrinsic} and \emph{tunable} magnetomechanical couplings allows for the construction of hybrid quantum devices with NV centers  and carbon-based nanostructures, as well as phonon-mediated quantum information
processing with spin qubits.

\end{abstract}
\maketitle

Carbon-based structures and devices are very commonly used in our everyday life and in state-of-the-art science and technology.
 In quantum information science, nitrogen-vacancy (NV) centers in diamond are outstanding solid state qubits due to their long coherence times and
high controllability \cite{SCI-314-281,nature-453-1043,natphys-2-408,prl-97-087601,PR-528-1}. In nano-mechanics, mechanical resonators made out of allotropes of carbon  (such as nanotubes \cite{nature-431-284,nl-6-2904,Nat-Nano-9-1007}, diamond \cite{Natcomm-5-3638,apl-103,apl-101},  and graphene \cite{SCI-315-490}),  are being extensively studied for fundamental research and practical applications \cite{Nat-Nano-8-165,prl-108-206811,prl-109-025503,nl-12-435,prl-112-133603,nl-14-2426,prl-113-047201,Nat-Nano-9-820,nl-14-2854,prl-112-223601,eprint-1504-08275}.

Recently, much attention has been paid to coupling NV spins in diamond to mechanical resonators, which can be achieved  extrinsically \cite{prb-79-041302,natphys-6-602,np-7-879,SCI-335-1603,nl-12,NC-6-8603,prappl-4-044003,pra-80-022335,pra-81-042323,pra-88-033614,prb-88-085201} or intrinsically \cite{prl-110-156402,prl-111-227602,prl-113-020503,Natcomm-5-4429,prb-88-064105}.
In the first case, the interaction arises from the relative
motion of the NV spin and a source of local magnetic field gradients \cite{prb-79-041302}. In such setups, a magnetic tip mounted on a vibrating cantilever \cite{prl-99-140403} is often used to generate the magnetic coupling between an NV spin and the mechanical motion \cite{prb-79-041302,natphys-6-602,np-7-879,SCI-335-1603,nl-12,NC-6-8603,prappl-4-044003,pra-80-022335,pra-81-042323,pra-88-033614,prb-88-085201}. However, creating very strong, well-controlled, local gradients remains challenging for such setups, in particular when arrays of NV centers are placed in close
proximity to the same cantilever. Thus far,  experiments with the extrinsic coupling scheme   have yet to reach the strong-coupling regime \cite{np-7-879,SCI-335-1603,nl-12,NC-6-8603}.
In the second case, the coupling of  a diamond cantilever to the spin of an embedded NV center  is induced by crystal strain during mechanical motion \cite{prl-110-156402,prl-111-227602,prl-113-020503,Natcomm-5-4429,prb-88-064105}. Unfortunately, the strain-induced interaction  between a single NV spin and the cantilever quantized motion  is inherently tiny \cite{prl-113-020503,Natcomm-5-4429}, which makes  the strong strain coupling at a single quantum level very challenging.

In this Letter, we propose that NV centers in diamond interfaced with carbon nanotubes can facilitate a  spin-nanomechanical hybrid device. This hybrid structure takes advantage of the unprecedented mechanical and electrical characteristics of  carbon nanotubes, as well as the exceptional coherence properties of  NV centers in diamond. We demonstrate that the physics of an NV center in diamond placed near a carbon nanotube with a dc current flowing through it can be well mapped to cavity quantum-electrodynamics (QED).  In particular, going beyond earlier work in this field \cite{prb-79-041302,natphys-6-602,np-7-879,SCI-335-1603,nl-12,NC-6-8603,prappl-4-044003,pra-80-022335,pra-81-042323,pra-88-033614,prb-88-085201,prl-110-156402,prl-111-227602,prl-113-020503,Natcomm-5-4429,prb-88-064105}, we show that the magnetomechanical interaction can be engineered and dynamically tuned by external control of the driving microwave fields and electric current through the nanotube. The resulting coupling strength can be roughly three orders of magnitude stronger than that for cold atoms coupled to nanowires  \cite{nl-12-435,prl-112-133603}.  An inherent advantage of our NV-nanotube hybrid system is the intrinsic nature of the coupling. Thus no additional components, such as external magnetic tips, are required to tune the coupling. Another distinct feature of this intrinsic coupling scheme is that it is scalable to  arrays of NV centers in diamond. This spin-nanomechanical structure with strong intrinsic magnetomechanical couplings would open up new avenues towards the design of hybrid quantum devices \cite{RMP-1} with NV centers and carbon-based nanostructures. It also allows for quantum information processing with NV spin qubits \cite{nature-514-72, prx-4-031022}, and  could serve as novel nanoscale sensors \cite{nature-455-644,nature-455-648,natphys-4-810,NC-5-4065,NC-6-6631} in physical and life science.

\emph {Model.--}
\begin{figure}[b]
\centerline{\includegraphics[bb=284 143 1058 396,totalheight=1.0in,clip]{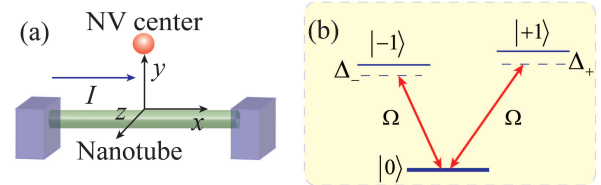}}
\caption{(Color online) (a) Schematic of a single NV center in a diamond nanocrystal located near a current-carrying nanotube.  (b) Level diagram of the driven NV center. }
\end{figure}
We consider a setup as shown in Fig. 1(a), where the magnetic field of a current-carrying nanotube
is coupled to an NV center spin embedded in a
nano-diamond. The nanotube of length $L$ is suspended along the $x$ axis at a distance $d$ from the
diamond nanocrystal. When it vibrates, $d$ varies by the nanotube's effective transverse displacement.
In the following, we assume the transverse displacement to be along the $y$ direction, and express it
with the oscillator operator $\hat{a}$ of the fundamental oscillating mode, i.e., $\hat{u}_y=(\hbar/2m\omega_\text{nt})^{1/2}(\hat{a}+\hat{a}^\dag)$, where
$m$ is the effective mass of the nanotube, and $\omega_\text{nt}$ is the mechanical vibration frequency \cite{Note1}.
The magnetic field $\vec{B}_{\text{nt}}(\vec{r})$ of the current-carrying nanotube at position $\vec{r}$
can be calculated by the  Biot-Savart law. For a long nanotube ($L\gg d$), the magnetic field has the form
$\vec{B}_{\text{nt}}(\vec{r})=\mu_0 I \vec{e}_x\times \vec{r}/2\pi|\vec{r}\vert^2$, in a reference frame with axes as in Fig. 1.
Here $\vec{e}_x$ is the unit vector in  the $x$ direction, and $I$ is the electric current  in the nanotube.

NV centers in diamond consist of a substitutional
nitrogen atom and an adjacent vacancy, which have a spin $S=1$
ground state, with zero-field splitting $D=2\pi\times 2.87$ GHz,
between the $\vert m_s=\pm1\rangle$ and $\vert m_s=0\rangle$ states.
For moderate applied magnetic fields, static and low frequency components of magnetic fields $B_z$ cause Zeeman
shifts of states $\vert m_s=\pm1\rangle$,  while external microwave  fields with magnetic field $\vec{B}_\text{dr}$ drive Rabi
oscillations between $\vert m_s=0\rangle$ and the excited states $\vert m_s=\pm1\rangle$, as shown
in Fig. 1(b). For convenience, we  denote as the $z$ axis the crystalline
axis of the NV center.

The interaction
of a single NV center located at $\vec{r}$
with  the  total magnetic field (external driving and from the nanotube) can be written as
$
\hat{H}_\text{NV}=\hbar D \hat{S}_z^2+\mu_Bg_s\ B_z\hat{S}_z+\mu_Bg_s (\vec{B}_{\text{nt}}(\vec{r})+ \vec{B}_\text{dr})\cdot\hat{\vec{S}}
$ with
$g_s=2$ the Land\'{e} factor of the NV center,  $\mu_B$ the Bohr magneton,  and $\hat{\vec{S}}$ the spin operator of the  NV center.
We consider using a single microwave field polarized in the $x$
direction, $\vec{B}_\text{dr}=B_\text{0}\cos\omega_0 t\vec{e}_x$, and place the NV center in the position where the
magnetic field of the nanotube is in the $z$ direction, $\vec{B}_\text{nt}=B_\text{nt}\vec{e}_z$.
The magnetic field $\vec{B}_{\text{nt}}(\vec{r})$ felt by the NV center is modulated by the nanotube's vibration.
Expanding the magnetic field $\vec{B}_{\text{nt}}(\vec{r})$ up to first order in $\hat{u}_y$, we have
$
\hat{H}_\text{NV}=\hbar D \hat{S}_z^2+\mu_Bg_s[\ B_z+B_\text{nt}(d)]\hat{S}_z+\mu_Bg_s   \vec{B}_\text{dr}\cdot\hat{\vec{S}}+\mu_Bg_s \hat{S}_z\partial_yB_\text{nt}\hat{u}_y.
$

We define $\hbar\Delta_\pm=\hbar D\pm\mu_Bg_s(B_z+B_\text{nt})-\hbar \omega_0$, $\hbar \Omega=\frac{\sqrt{2}}{4}\mu_Bg_sB_0$, and restrict the following discussion to symmetric detunings, $\Delta_+=\Delta_-=\Delta$.
When $\vert\Delta\vert\gg\Omega$, we  obtain the effective Hamiltonian \cite{Note1}
\begin{eqnarray}
\hat{\mathcal{H}}_q&=&\hbar \omega_\text{nt} \hat{a}^\dag\hat{a}+\frac{1}{2}\hbar \Lambda \hat{\sigma}_z+\hbar g(  \hat{\sigma}_++\hat{\sigma}_-)(\hat{a}^\dag+\hat{a}).
\end{eqnarray}
Here, $\Lambda=2\Omega^2/\Delta$, $\hbar g=\mu_Bg_s(\hbar/2m\omega_\text{nt})^{1/2}\partial_yB_\text{nt}$, and we switch to the new basis $\{\vert \mathcal {B}\rangle=\frac{1}{\sqrt{2}}(\vert +1\rangle+\vert-1\rangle),\vert \mathcal {D}\rangle=\frac{1}{\sqrt{2}}(\vert +1\rangle-\vert-1\rangle)\}$, with $\hat{\sigma}_z=\vert \mathcal {B}\rangle\langle \mathcal {B}\vert-\vert \mathcal {D}\rangle\langle \mathcal {D}\vert$, $\hat{\sigma}_+=\vert \mathcal {B}\rangle\langle \mathcal {D}\vert$, and $\hat{\sigma}_-=\vert \mathcal {D}\rangle\langle \mathcal {B}\vert$.
The states $\vert \mathcal {B}\rangle$ and $\vert \mathcal {D}\rangle$ are often referred to bright and dark states for NV spins \cite{prb-79-041302,prl-114-120501}.
If we choose $\Lambda\simeq\omega_\text{nt}$, then, under the rotating-wave approximation we
obtain the standard Jaynes-Cummings (JC) Hamiltonian
\begin{eqnarray}\label{H}
\hat{\mathcal{H}}_\text{JC}&=&\hbar \omega_\text{nt} \hat{a}^\dag\hat{a}+\frac{1}{2}\hbar \Lambda\hat{ \sigma}_z+\hbar g(  \hat{\sigma}_+ \hat{a}+\hat{\sigma}_- \hat{a}^\dag).
\end{eqnarray}
However, if we choose $\Lambda\simeq-\omega_\text{nt}$, which can be controlled by the parameters $\Delta$ and $\Omega$,
we obtain the Anti-Jaynes-Cummings (AJC) Hamiltonian
\begin{eqnarray}\label{H2}
\hat{\mathcal{H}}_\text{AJC}&=&\hbar \omega_\text{nt} \hat{a}^\dag\hat{a}+\frac{1}{2}\hbar \Lambda\hat{ \sigma}_z+\hbar g(  \hat{\sigma}_+ \hat{a}^\dag+\hat{\sigma}_- \hat{a}).
\end{eqnarray}

Thus, our system mimics the standard  model in cavity QED of a single atom coupled to a single cavity mode.
The type of  interactions can be designed by external control over the driving fields.
This mapping
allows the powerful toolbox of cavity QED to be transferred to
these systems.

\emph  {Two proposed experimental setups.--}
\begin{figure}[b]
\centerline{\includegraphics[bb=0 22 847 268,totalheight=1in,clip]{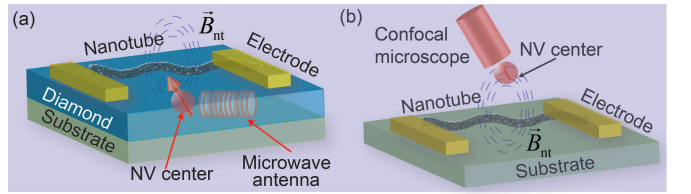}}
\caption{(Color online) Schematic of the nanotube-NV hybrid setup. (a) A current-carrying nanotube is suspended above a diamond sample, in which
individual optically-resolvable NV centers are implanted $5-10$ nm below its surface.  (b) A single NV defect hosted in a diamond nanocrystal with a size about 10 nm  is positioned near the nanotube.}
\end{figure}
We consider  two different   designs for coupling an NV spin to a current-carrying carbon nanotube resonator. Fig. 2 (a) displays a nanotube, carrying a dc
current,  suspended above a bulk single crystal diamond sample. Individual, optically-resolvable NV centers are implanted $5-10$ nm below the surface of the
diamond sample \cite{SCI-335-1603,nl-13-3733}.  Fig. 2 (b) shows another feasible design, where a diamond nanoparticle  hosting a single NV center  is closely
placed near the nanotube.  Diamond nanoparticles can have a size of less than 10 nanometers,
and only host one NV defect \cite{np-7-879}. The spin states of  NV centers can be controlled and manipulated by microwaves from external microwave  antennas.  A confocal microscope can be used to excite and polarize the NV spin, and detect  photoluminescence to read out the NV spin polarization.

Carbon nanotubes can possess current-carrying capacities
exceeding 10 $\mu$A/$\text{nm}^2$ \cite{prl-84-2941,prl-86-3128,prl-92-106804,SCI-272-523,SCI-280-1744,nl-12-1603}, with lengths which can range from tens of nanometers to tens of micrometers. In experiments, the carbon nanotube could be actuated and deflected electrostatically over several nanometers with AC and DC voltages applied to the gate electrodes. Thus, the distance between the nanotube and the NV center can be fine-tuned electrostatically. A recent experiment \cite{eprint-1504-08275},  has reported  a device  with a	graphene membrane	 suspended some $10-50$	 nm	above a single NV center.

To evaluate the single spin-phonon coupling strength $g$, we use
the magnetic field  generated by an infinite long tube, as given by the Biot-Savart law.
In this case, it reads
\begin{eqnarray}
g&=&\frac{\mu_Bg_s \mu_0 I}{2\pi\sqrt{2\hbar m \omega_\text{nt}}d^2}.
\end{eqnarray}
This magnetomechanical coupling strength depends on the the dimensions of the nanotube and the distance $d$, as well as the current $I$ flowing through the nanotube. Thus, it can be easily tuned by control of the system parameters.

We consider a carbon nanotube of length $L\sim2$ $\mu\text{m}$, radius $r\sim1.5$ $\text{nm}$, and wall thickness $t\sim0.335$ $\text{nm}$, suspended at a distance $d\sim30$ $\text{nm}$ from the NV center \cite{Note1}. The tube carries a dc current $I\sim60$ $\mu \text{A}$, and vibrates at a frequency
$\omega_\text{nt}/2\pi\sim2  $ MHz, with effective mass $m\sim7\times10^{-21}$ kg \cite{Note1}.
With the above given parameters, one can obtain $g/2\pi\sim10$ kHz.
By changing the distance $d$ and the dimensions of the nanotube, as well as the current $I$ flowing through the nanotube, the coupling strength $g$ can be further adjusted (see Fig. 3). For a much closer distance $d\sim 10$ nm, the magnetomechanical coupling strength can even reach $g/2\pi\sim100$ kHz.
This coupling strength is comparable to that of a single NV spin coupled to a vibrating cantilever with a strong local magnet \cite{prb-79-041302,natphys-6-602} or a superconducting circuit \cite{prl-105-210501,prb-81-033614,nature-478-221,prl-105-140502,prb-87-144516,pra-88-012329}, and is  about a factor of 1000
larger than the coupling achieved with cold atoms \cite{nl-12-435,prl-112-133603}.

\begin{figure}[t]
\centerline{\includegraphics[bb=26 199 301 379,totalheight=1.35in,clip]{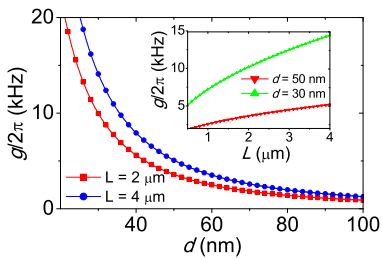}}
\caption{(Color online) Single spin-phonon coupling strength versus the distance between the NV center and the nanotube.  The relevant parameters here
are $r\sim1.5$ $\text{nm}$, $t\sim0.335$ $\text{nm}$, $I\sim60$ $\mu \text{A}$.
In the inset  the coupling strength versus the
length of the nanotube.  }
\end{figure}

 \emph {Dephasing and dissipation.--}
In realistic situations, we need to
consider spin dephasing and mechanical dissipation. The full dynamics of our
system that takes these incoherent processes into account
is described by the master equation
\begin{eqnarray}
\label{M1}
\frac{d\hat{\rho}(t)}{dt}&=&-\frac{i}{\hbar}[\hat{\mathcal{H}}_\text{JC},\hat{\rho}]+\gamma_\text{s}\mathcal{D}[\hat{\sigma}_z]\hat{\rho}\nonumber\\
&&+n_\text{th}\gamma_\text{m}\mathcal{D}[\hat{a}^\dag]\hat{\rho}+(n_\text{th}+1)\gamma_\text{m}\mathcal{D}[\hat{a}]\hat{\rho}
\end{eqnarray}
with $\mathcal{D}[\hat{o}]\hat{\rho}=\hat{o}\hat{\rho}\hat{o}^\dag-\frac{1}{2}\hat{o}^\dag\hat{o}\hat{\rho}-\frac{1}{2}\hat{\rho}\hat{o}^\dag\hat{o}$
for a given operator $\hat{o}$.
The strong coupling regime can be reached if the coherent coupling strength $g$ exceeds
both the electronic spin decay rate $\gamma_s$  and the
intrinsic damping rate of the mechanical mode $\gamma_m$, i.e., $g> \{ \gamma_s,n_\text{th}\gamma_m\}$, with
$n_\text{th}=(e^{\hbar \omega_\text{nt}/k_\text{B}T}-1)^{-1}$ the thermal phonon number at the environment
temperature $T$.
For a mechanical resonator with frequency $\omega_\text{m}$ and quality factor $Q$, the
mechanical damping rate is $\gamma_\text{m}=\omega_\text{m}/Q$.
The recent fabrication of carbon nanotube resonators can possess  quality factors exceeding $10^5$ \cite{Nat-Nano-9-1007}.
Together with the oscillator frequency
$\omega_\text{nt}/2\pi\sim 2$ MHz \cite{Note1}, this value of $Q$ implies an oscillator
damping rate   $\gamma_\text{m}/ 2\pi\sim 20$ Hz, and would translate into phonon mean free path $l_c\sim QL \sim 10$ cm \cite{Note1}. Assuming an environmental temperature $T\sim 10 $ $\text{mK}$ in a dilution refrigerator, the thermal phonon
number is about $n_\text{th}\sim 100$. Therefore, we obtain $g>  n_\text{th}\gamma_m$.
When it comes to the
NV center, the dephasing time $T_2$ can be increased to several
milliseconds in ultrapure diamond \cite{Naure-Mat}, leading to a dephasing rate $\gamma_\text{s}/2\pi\sim1$ kHz. We can ignore
single spin relaxation, as  $T_1$ can be several minutes at low
temperatures. Therefore, the strong-coupling regime can be reached in this setup, i.e., $g> \{ \gamma_s,n_\text{th}\gamma_m\}$.

The strong-coupling regime described by the Hamiltonian (\ref{H}) enables coherent quantum state transfer between the spin and the
resonator. Moreover, in combination with optical pumping and detection techniques for spin qubits, this would provide the basic ingredients
for detecting and manipulating  quantum states of the nanotube resonator.

\begin{figure}[b]
\centerline{\includegraphics[bb=22  252 320 377,totalheight=1.35in,clip]{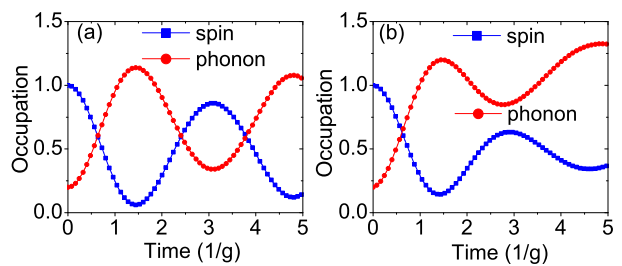}}
\caption{(Color online) (a) Vacuum Rabi oscillations of  an NV spin  coupled to a nanotube mechanical resonator without dissipations. The
initial state of the NV spin is $\vert \mathcal{B}\rangle$.
(b) Same as (a) but with
dissipations for the spin and the mechanical resonator. The relevant parameters here are chosen as
$n_\text{th}\sim100$, $\gamma_\text{m}\sim10^{-3}g$, and $\gamma_\text{s} \sim0.1g$.}
\end{figure}
In Fig. 4, we show the numerical simulations of quantum dynamics of the coupled system through
solving the master equation (\ref{M1}). As the initial state, we take the product state of the NV
spin  and the mechanical resonator with the occupation number $n_\text{m}=0.2$, e.g., as a result of side-band cooling \cite{nature-475-359,nature-478-89}.
In the time domain, vacuum Rabi oscillations are a direct
evidence of the coherent energy exchange between the
spin qubit and the resonator phonon mode.
We obtain  numerical results for the time evolution
of the mean phonon number and the probability for the NV center being in the
excited state. We find that vacuum Rabi oscillations   can occur for these parameters.
In such a process, the spin state can be
transferred from the NV center to the nanotube resonator, and vice versa.

 \emph {Elementary quantum information.--}
The magnetomechanical interaction allows us to use the nanotube vibration mode as a quantum  bus to perform more complex tasks.
Controlled spin-spin couplings could be realized for two distant NV centers separated by micrometer distances. Based on these
effective interactions, we now explore the possibility of implementing quantum information processing with spin qubits.

We consider two separated NV centers coupled to the same vibration mode of the nanotube in the dispersive regime $|\Lambda- \omega_\text{nt}|\gg g$. This will lead to an effective long range spin-spin interaction via the exchange of virtual phonons \cite{Note1},
$
\hat{\mathcal{H}}_\text{s-s}= \hbar \lambda_\text{eff} (\hat{\sigma}_+^1\hat{\sigma}_-^2+\hat{\sigma}_-^1\hat{\sigma}_+^2),
$
with the coupling strength $\lambda_\text{eff}=g^2/|\Lambda- \omega_\text{nt}|$.
The coherence length of the phonon mediated
NV spin coupling is about $l_{c}\sim QL$ \cite{Note1}, which can be much larger than the distance between two NV spins separated by a
distance on the order of the nanotube's length.
If we choose  $|\Lambda- \omega_\text{nt}|/2\pi\sim 1$ MHz, and $g/2\pi\sim100$ kHz, then we can obtain the spin-spin coupling strength $\lambda_\text{eff}/2\pi\sim 10$ kHz. This strong spin-spin interaction allows
for the implementation of a SWAP gate and quantum states transfer between two NV centers.

In the following, we encode the $j$th \emph{logical} qubit  in the two spin states of the $j$th NV center,  i.e.,  $\vert0\rangle_q^j=\vert 0\rangle_j$ and $\vert1\rangle_q^j=\vert \mathcal {D}\rangle_j$. Such qubit encoding has proven to be more robust against  low-frequency
magnetic-field noise \cite{prl-114-120501}.  Our main task is to realize a SWAP gate and quantum information transfer between two qubits.
To implement this protocol, we need a microwave to drive  the transition between the qubit state $\vert 0\rangle_j$ and the bright state $\vert\mathcal {B}\rangle_j$ in each qubit with Rabi frequency $\Omega_j$ and frequency detuning $\delta_j$. The dynamics of the entire system is described by
$
\hat{\mathcal{H}}=\sum_j\hbar \delta_j \vert \mathcal{B}\rangle_{jj}\langle \mathcal{B}\vert+\sum_j[{\hbar \Omega_j\vert \mathcal{B}\rangle_{jj}\langle 0\vert +\text{H.c.}}]+\hat{\mathcal{H}}_\text{s-s}.
$
 The spin-spin interaction can be diagonalized with the states $\vert \pm\rangle_q=1/\sqrt{2}[\vert \mathcal{B}\rangle_1\vert\mathcal{D}\rangle_2\pm\vert\mathcal{D}\rangle_1\vert \mathcal{B}\rangle_2]$.
\begin{figure}[b]
\centerline{\includegraphics[bb=22  167 345 388,totalheight=2in,clip]{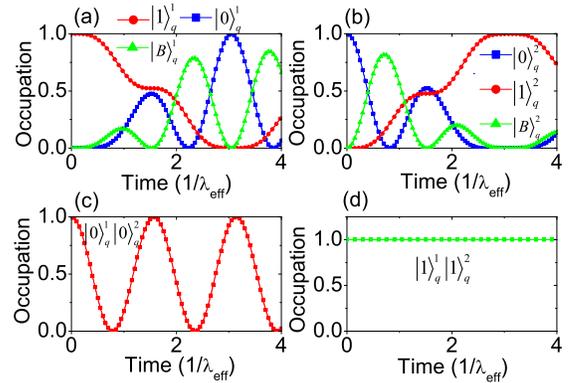}}
\caption{(Color online) (a)  Time evolution of the probabilities for the first NV center being in the states $\vert 1\rangle,\vert0\rangle$, and $\vert\mathcal{B}\rangle$. (b) Same as in (a) but for the second NV center. In both cases, the initial state is chosen as $\vert 1\rangle_q^1\vert0\rangle_q^2$.  (c) and (d):  Time evolution of the probabilities for the two NV spins in the states $\vert0\rangle^1_q\vert0\rangle^2_q$, and $\vert1\rangle^1_q\vert1\rangle^2_q$ respectively. The relevant parameters here are $\Omega_1\simeq\Omega_2= 2\lambda_\text{eff}$, and $ \delta_1=\delta_2=0$.
}
\end{figure}
It is easy to show that in the subspace defined by $\{\vert0,1\rangle_q\equiv \vert0\rangle_q^1\vert1\rangle^2_q,\vert+\rangle_q,\vert-\rangle_q,  \vert1,0\rangle_q\equiv\vert1\rangle_q^1\vert0\rangle^2_q \}$, the system Hamiltonian has the form \cite{Note1}
\begin{eqnarray}\label{H3}
  \hat{\mathcal{H}} &=& \hbar\delta_+\vert+\rangle_{qq}\langle+\vert-\hbar\delta_-\vert-\rangle_{qq}\langle-\vert +\hbar\bar{\Omega}_1\vert+\rangle_{qq}\langle0,1\vert \nonumber \\
  &&+\hbar\bar{\Omega}_1\vert-\rangle_{qq}\langle0,1\vert+\hbar\bar{\Omega}_2\vert+\rangle_{qq}\langle1,0\vert \nonumber\\
  &&-\hbar\bar{\Omega}_2\vert-\rangle_{qq}\langle1,0\vert+\text{H.c.}
\end{eqnarray}
with $\delta_+=\lambda_\text{eff}+\frac{\delta_1+\delta_2}{2}$, $\delta_-=\lambda_\text{eff}-\frac{\delta_1+\delta_2}{2}$, $\bar{\Omega}_{j}=\Omega_{j}/\sqrt{2},j=1,2$.  Thus we can find that if the two qubits are initially prepared in the state $\vert 0\rangle_q^1\vert1\rangle_q^2$ or $\vert 1\rangle_q^1\vert0\rangle_q^2$, then the dynamics of the system will be confined in the subspace governed by the Hamiltonian (\ref{H3}). In Fig. 5 we present detailed numerical simulations for the dynamics of the coupled system.
It can be found that at the moment $T_\text{sw}=\pi/\lambda_\text{eff}$, the system evolves from the state $\vert0\rangle_q^1\vert1\rangle^2_q$
to the state $\vert 1\rangle_q^1\vert0\rangle^2_q$ via the intermediate states $\vert \pm\rangle_q $ through microwave driving, and vice versa.
The state $\vert 1\rangle_q^1\vert 1\rangle^2_q$ remains unchanged during this process, while the system  will be
brought from the state $\vert 0\rangle_q^1\vert 0\rangle^2_q$ to the state $\vert \mathcal{B}\rangle_1\vert 0\rangle^2_q$ or $\vert 0\rangle_q^1\vert \mathcal{B}\rangle_2$, and back again at the moment $T_\text{sw}$.
Therefore, in the language of quantum information theory, this operation corresponds to a SWAP  gate.
This gate can be exploited to realize quantum state transfer between two NV spins, i.e.,
$
(\alpha \vert0\rangle_q^1+\beta\vert1\rangle_q^1)\vert1\rangle_q^2\rightarrow\vert1\rangle_q^1(\alpha \vert0\rangle_q^2+\beta\vert1\rangle_q^2).
$

The gate fidelity is limited by the following factors: (i) spin decoherence induced by phonon excitations with an effective decay rate $\Gamma\simeq n_\text{th}\gamma_\text{m}g^2/|\Lambda-\omega_\text{nt}|^2$; (ii) single spin dephasing due to low-frequency noise with a dephasing rate $\gamma_\text{s}$, which is assumed to be Markovian for simplicity. Recent work has shown that by coupling a single NV spin to another two-level system and encoding quantum information in the dark-states $\vert0\rangle$ and $\vert \mathcal{D}\rangle$, the  coherent time $T_2$ can be prolonged \cite{prl-114-120501}. For
isotopically purified diamond, we can safely choose $T_2\sim 1$ ms.  Taking these factors together, we find the gate fidelity can be estimated as $F\sim (1-T_\text{sw}\Gamma-T_\text{sw}/T_2)>0.95$ for the given parameters, with an operating time $T_\text{sw}\sim 50$ $\mu s$.

\emph {Conclusions.--}
We have proposed a spin-nanomechanical hybrid device where  a single NV center spin in  diamond  is coupled to the vibrational mode of a suspended current-carrying carbon nanotube. It makes the strong spin-mechanical coupling at a single quantum level feasible, and allows fast mechanical control of spin qubits and efficient phonon
cooling by an NV center.
Such a device can find applications in phonon-mediated  quantum information processing
with NV spin qubits, and could serve as novel
nanoscale sensors for detecting tiny pressure, temperature, electric and magnetic field changes.

PBL thanks Hongyan  Li, Jie Liu, Zhou Li, and  Minggang Xia for valuable discussions. PBL is supported by the NSFC under
 Grant No. 11474227.  ZLX and PR acknowledge support through  the Austrian Science Fund (FWF) through SFB FOQUS, and the START grant Y 591-N16 and the European
Commission through Marie Sklodowska-Curie Grant
No. IF 657788. FN  is partially supported by the
RIKEN iTHES Project,
the MURI Center for Dynamic Magneto-Optics
via the AFOSR award number FA9550-14-1-0040,
the IMPACT program of JST,
and a Grant-in-Aid for Scientific Research (A).
F. N. also acknowledges the support of a grant from the John Templeton Foundation.
Part of the
simulations are coded in PYTHON using the QUTIP library \cite{CPC}.






\begin{thebibliography}{69}%
\makeatletter
\providecommand \@ifxundefined [1]{%
 \@ifx{#1\undefined}
}%
\providecommand \@ifnum [1]{%
 \ifnum #1\expandafter \@firstoftwo
 \else \expandafter \@secondoftwo
 \fi
}%
\providecommand \@ifx [1]{%
 \ifx #1\expandafter \@firstoftwo
 \else \expandafter \@secondoftwo
 \fi
}%
\providecommand \natexlab [1]{#1}%
\providecommand \enquote  [1]{``#1''}%
\providecommand \bibnamefont  [1]{#1}%
\providecommand \bibfnamefont [1]{#1}%
\providecommand \citenamefont [1]{#1}%
\providecommand \href@noop [0]{\@secondoftwo}%
\providecommand \href [0]{\begingroup \@sanitize@url \@href}%
\providecommand \@href[1]{\@@startlink{#1}\@@href}%
\providecommand \@@href[1]{\endgroup#1\@@endlink}%
\providecommand \@sanitize@url [0]{\catcode `\\12\catcode `\$12\catcode
  `\&12\catcode `\#12\catcode `\^12\catcode `\_12\catcode `\%12\relax}%
\providecommand \@@startlink[1]{}%
\providecommand \@@endlink[0]{}%
\providecommand \url  [0]{\begingroup\@sanitize@url \@url }%
\providecommand \@url [1]{\endgroup\@href {#1}{\urlprefix }}%
\providecommand \urlprefix  [0]{URL }%
\providecommand \Eprint [0]{\href }%
\providecommand \doibase [0]{http://dx.doi.org/}%
\providecommand \selectlanguage [0]{\@gobble}%
\providecommand \bibinfo  [0]{\@secondoftwo}%
\providecommand \bibfield  [0]{\@secondoftwo}%
\providecommand \translation [1]{[#1]}%
\providecommand \BibitemOpen [0]{}%
\providecommand \bibitemStop [0]{}%
\providecommand \bibitemNoStop [0]{.\EOS\space}%
\providecommand \EOS [0]{\spacefactor3000\relax}%
\providecommand \BibitemShut  [1]{\csname bibitem#1\endcsname}%
\let\auto@bib@innerbib\@empty
\bibitem [{\citenamefont {Childress}\ \emph {et~al.}(2006)\citenamefont
  {Childress}, \citenamefont {Dutt}, \citenamefont {Taylor}, \citenamefont
  {Zibrov}, \citenamefont {Jelezko}, \citenamefont {Wrachtrup}, \citenamefont
  {Hemmer},\ and\ \citenamefont {Lukin}}]{SCI-314-281}%
  \BibitemOpen
  \bibfield  {author} {\bibinfo {author} {\bibfnamefont {L.}~\bibnamefont
  {Childress}}, \bibinfo {author} {\bibfnamefont {M.~V.~Gurudev}\ \bibnamefont
  {Dutt}}, \bibinfo {author} {\bibfnamefont {J.~M.}\ \bibnamefont {Taylor}},
  \bibinfo {author} {\bibfnamefont {A.~S.}\ \bibnamefont {Zibrov}}, \bibinfo
  {author} {\bibfnamefont {F.}~\bibnamefont {Jelezko}}, \bibinfo {author}
  {\bibfnamefont {J.}~\bibnamefont {Wrachtrup}}, \bibinfo {author}
  {\bibfnamefont {P.~R.}\ \bibnamefont {Hemmer}}, \ and\ \bibinfo {author}
  {\bibfnamefont {M.~D.}\ \bibnamefont {Lukin}},\ }\bibfield  {title} {\enquote
  {\bibinfo {title} {Coherent dynamics of coupled electron and nuclear spin
  qubits in diamond},}\ }\href@noop {} {\bibfield  {journal} {\bibinfo
  {journal} {Science}\ }\textbf {\bibinfo {volume} {314}},\ \bibinfo {pages}
  {281} (\bibinfo {year} {2006})}\BibitemShut {NoStop}%
\bibitem [{\citenamefont {Hanson}\ and\ \citenamefont
  {Awschalom}(2008)}]{nature-453-1043}%
  \BibitemOpen
  \bibfield  {author} {\bibinfo {author} {\bibfnamefont {Ronald}\ \bibnamefont
  {Hanson}}\ and\ \bibinfo {author} {\bibfnamefont {David~D.}\ \bibnamefont
  {Awschalom}},\ }\bibfield  {title} {\enquote {\bibinfo {title} {Coherent
  manipulation of single spins in semiconductors},}\ }\href@noop {} {\bibfield
  {journal} {\bibinfo  {journal} {Nature (London)}\ }\textbf {\bibinfo {volume}
  {453}},\ \bibinfo {pages} {1043} (\bibinfo {year} {2008})}\BibitemShut
  {NoStop}%
\bibitem [{\citenamefont {Gaebel}\ \emph {et~al.}(2006)\citenamefont {Gaebel},
  \citenamefont {Domhan}, \citenamefont {Popa}, \citenamefont {Wittmann},
  \citenamefont {Neumann}, \citenamefont {Jelezko}, \citenamefont {Rabeau},
  \citenamefont {Stavrias}, \citenamefont {Greentree}, \citenamefont {Prawer},
  \citenamefont {Meijer}, \citenamefont {Twamley}, \citenamefont {Hemmer},\
  and\ \citenamefont {Wrachtrup}}]{natphys-2-408}%
  \BibitemOpen
  \bibfield  {author} {\bibinfo {author} {\bibfnamefont {Torsten}\ \bibnamefont
  {Gaebel}}, \bibinfo {author} {\bibfnamefont {Michael}\ \bibnamefont
  {Domhan}}, \bibinfo {author} {\bibfnamefont {Iulian}\ \bibnamefont {Popa}},
  \bibinfo {author} {\bibfnamefont {Christoffer}\ \bibnamefont {Wittmann}},
  \bibinfo {author} {\bibfnamefont {Philipp}\ \bibnamefont {Neumann}}, \bibinfo
  {author} {\bibfnamefont {Fedor}\ \bibnamefont {Jelezko}}, \bibinfo {author}
  {\bibfnamefont {James~R.}\ \bibnamefont {Rabeau}}, \bibinfo {author}
  {\bibfnamefont {Nikolas}\ \bibnamefont {Stavrias}}, \bibinfo {author}
  {\bibfnamefont {Andrew~D.}\ \bibnamefont {Greentree}}, \bibinfo {author}
  {\bibfnamefont {Steven}\ \bibnamefont {Prawer}}, \bibinfo {author}
  {\bibfnamefont {Jan}\ \bibnamefont {Meijer}}, \bibinfo {author}
  {\bibfnamefont {Jason}\ \bibnamefont {Twamley}}, \bibinfo {author}
  {\bibfnamefont {Philip~R.}\ \bibnamefont {Hemmer}}, \ and\ \bibinfo {author}
  {\bibfnamefont {J\"{o}rg}\ \bibnamefont {Wrachtrup}},\ }\bibfield  {title}
  {\enquote {\bibinfo {title} {Room-temperature coherent coupling of single
  spins in diamond},}\ }\href@noop {} {\bibfield  {journal} {\bibinfo
  {journal} {Nat. Phys.}\ }\textbf {\bibinfo {volume} {2}},\ \bibinfo {pages}
  {408} (\bibinfo {year} {2006})}\BibitemShut {NoStop}%
\bibitem [{\citenamefont {Hanson}\ \emph {et~al.}(2006)\citenamefont {Hanson},
  \citenamefont {Mendoza}, \citenamefont {Epstein},\ and\ \citenamefont
  {Awschalom}}]{prl-97-087601}%
  \BibitemOpen
  \bibfield  {author} {\bibinfo {author} {\bibfnamefont {R.}~\bibnamefont
  {Hanson}}, \bibinfo {author} {\bibfnamefont {F.~M.}\ \bibnamefont {Mendoza}},
  \bibinfo {author} {\bibfnamefont {R.~J.}\ \bibnamefont {Epstein}}, \ and\
  \bibinfo {author} {\bibfnamefont {D.~D.}\ \bibnamefont {Awschalom}},\
  }\bibfield  {title} {\enquote {\bibinfo {title} {Polarization and readout of
  coupled single spins in diamond},}\ }\href@noop {} {\bibfield  {journal}
  {\bibinfo  {journal} {Phys.\ Rev. Lett.}\ }\textbf {\bibinfo {volume} {97}},\
  \bibinfo {pages} {087601} (\bibinfo {year} {2006})}\BibitemShut {NoStop}%
\bibitem [{\citenamefont {Doherty}\ \emph {et~al.}(2013)\citenamefont
  {Doherty}, \citenamefont {Manson}, \citenamefont {Delaney}, \citenamefont
  {Jelezko}, \citenamefont {Wrachtrup},\ and\ \citenamefont
  {Hollenberg}}]{PR-528-1}%
  \BibitemOpen
  \bibfield  {author} {\bibinfo {author} {\bibfnamefont {Marcus~W.}\
  \bibnamefont {Doherty}}, \bibinfo {author} {\bibfnamefont {Neil~B.}\
  \bibnamefont {Manson}}, \bibinfo {author} {\bibfnamefont {Paul}\ \bibnamefont
  {Delaney}}, \bibinfo {author} {\bibfnamefont {Fedor}\ \bibnamefont
  {Jelezko}}, \bibinfo {author} {\bibfnamefont {J\"{o}rg}\ \bibnamefont
  {Wrachtrup}}, \ and\ \bibinfo {author} {\bibfnamefont {Lloyd~C.L.}\
  \bibnamefont {Hollenberg}},\ }\bibfield  {title} {\enquote {\bibinfo {title}
  {The nitrogen-vacancy colour centre in diamond},}\ }\href@noop {} {\bibfield
  {journal} {\bibinfo  {journal} {Phys. Rep.}\ }\textbf {\bibinfo {volume}
  {528}},\ \bibinfo {pages} {1} (\bibinfo {year} {2013})}\BibitemShut {NoStop}%
\bibitem [{\citenamefont {Sazonova}\ \emph {et~al.}(2004)\citenamefont
  {Sazonova}, \citenamefont {Yaish}, \citenamefont {Ustunel}, \citenamefont
  {Roundy}, \citenamefont {Arias},\ and\ \citenamefont
  {McEuen}}]{nature-431-284}%
  \BibitemOpen
  \bibfield  {author} {\bibinfo {author} {\bibfnamefont {Vera}\ \bibnamefont
  {Sazonova}}, \bibinfo {author} {\bibfnamefont {Yuval}\ \bibnamefont {Yaish}},
  \bibinfo {author} {\bibfnamefont {Hande}\ \bibnamefont {\"{U}st\"{u}nel}}, \bibinfo
  {author} {\bibfnamefont {David}\ \bibnamefont {Roundy}}, \bibinfo {author}
  {\bibfnamefont {Tom\'{a}s~A.}\ \bibnamefont {Arias}}, \ and\ \bibinfo
  {author} {\bibfnamefont {Paul~L.}\ \bibnamefont {McEuen}},\ }\bibfield
  {title} {\enquote {\bibinfo {title} {A tunable carbon nanotube
  electromechanical oscillator},}\ }\href@noop {} {\bibfield  {journal}
  {\bibinfo  {journal} {Nature}\ }\textbf {\bibinfo {volume} {431}},\ \bibinfo
  {pages} {284} (\bibinfo {year} {2004})}\BibitemShut {NoStop}%
\bibitem [{\citenamefont {Witkamp}\ \emph {et~al.}(2006)\citenamefont
  {Witkamp}, \citenamefont {Poot},\ and\ \citenamefont {van~der
  Zant}}]{nl-6-2904}%
  \BibitemOpen
  \bibfield  {author} {\bibinfo {author} {\bibfnamefont {Benoit}\ \bibnamefont
  {Witkamp}}, \bibinfo {author} {\bibfnamefont {Menno}\ \bibnamefont {Poot}}, \
  and\ \bibinfo {author} {\bibfnamefont {Herre S.~J.}\ \bibnamefont {van~der
  Zant}},\ }\bibfield  {title} {\enquote {\bibinfo {title} {Bending-mode
  vibration of a suspended nanotube resonator},}\ }\href@noop {} {\bibfield
  {journal} {\bibinfo  {journal} {Nano Lett.}\ }\textbf {\bibinfo {volume}
  {6}},\ \bibinfo {pages} {2904} (\bibinfo {year} {2006})}\BibitemShut
  {NoStop}%
\bibitem [{\citenamefont {Moser}\ \emph {et~al.}(2014)\citenamefont {Moser},
  \citenamefont {Eichler}, \citenamefont {Guttinger}, \citenamefont {Dykman},\
  and\ \citenamefont {Bachtold}}]{Nat-Nano-9-1007}%
  \BibitemOpen
  \bibfield  {author} {\bibinfo {author} {\bibfnamefont {J.}~\bibnamefont
  {Moser}}, \bibinfo {author} {\bibfnamefont {A.}~\bibnamefont {Eichler}},
  \bibinfo {author} {\bibfnamefont {J.}~\bibnamefont {G\"{u}ttinger}}, \bibinfo
  {author} {\bibfnamefont {M.~I.}\ \bibnamefont {Dykman}}, \ and\ \bibinfo
  {author} {\bibfnamefont {A.}~\bibnamefont {Bachtold}},\ }\bibfield  {title}
  {\enquote {\bibinfo {title} {Nanotube mechanical resonators with quality
  factors of up to 5 million},}\ }\href@noop {} {\bibfield  {journal} {\bibinfo
   {journal} {Nat. Nanotech.}\ }\textbf {\bibinfo {volume} {9}},\ \bibinfo
  {pages} {1007} (\bibinfo {year} {2014})}\BibitemShut {NoStop}%
\bibitem [{\citenamefont {Tao}\ \emph {et~al.}(2014)\citenamefont {Tao},
  \citenamefont {Boss}, \citenamefont {Moores},\ and\ \citenamefont
  {Degen}}]{Natcomm-5-3638}%
  \BibitemOpen
  \bibfield  {author} {\bibinfo {author} {\bibfnamefont {Y.}~\bibnamefont
  {Tao}}, \bibinfo {author} {\bibfnamefont {J.~M.}\ \bibnamefont {Boss}},
  \bibinfo {author} {\bibfnamefont {B.~A.}\ \bibnamefont {Moores}}, \ and\
  \bibinfo {author} {\bibfnamefont {C.~L.}\ \bibnamefont {Degen}},\ }\bibfield
  {title} {\enquote {\bibinfo {title} {Single-crystal diamond nanomechanical
  resonators with quality factors exceeding one million},}\ }\href@noop {}
  {\bibfield  {journal} {\bibinfo  {journal} {Nat. Commun.}\ }\textbf {\bibinfo
  {volume} {5}},\ \bibinfo {pages} {3638} (\bibinfo {year} {2014})}\BibitemShut
  {NoStop}%
\bibitem [{\citenamefont {Burek}\ \emph {et~al.}(2013)\citenamefont {Burek},
  \citenamefont {Ramos}, \citenamefont {Patel}, \citenamefont {Frank},\ and\
  \citenamefont {Loncar}}]{apl-103}%
  \BibitemOpen
  \bibfield  {author} {\bibinfo {author} {\bibfnamefont {Michael~J.}\
  \bibnamefont {Burek}}, \bibinfo {author} {\bibfnamefont {Daniel}\
  \bibnamefont {Ramos}}, \bibinfo {author} {\bibfnamefont {Parth}\ \bibnamefont
  {Patel}}, \bibinfo {author} {\bibfnamefont {Ian~W.}\ \bibnamefont {Frank}}, \
  and\ \bibinfo {author} {\bibfnamefont {Marko}\ \bibnamefont {Loncar}},\
  }\bibfield  {title} {\enquote {\bibinfo {title} {Nanomechanical resonant
  structures in single-crystal diamond},}\ }\href@noop {} {\bibfield  {journal}
  {\bibinfo  {journal} {Appl. Phys. Lett.}\ }\textbf {\bibinfo {volume}
  {103}},\ \bibinfo {pages} {131904} (\bibinfo {year} {2013})}\BibitemShut
  {NoStop}%
\bibitem [{\citenamefont {Ovartchaiyapong}\ \emph {et~al.}(2012)\citenamefont
  {Ovartchaiyapong}, \citenamefont {Pascal}, \citenamefont {Myers},
  \citenamefont {Lauria},\ and\ \citenamefont {Jayich}}]{apl-101}%
  \BibitemOpen
  \bibfield  {author} {\bibinfo {author} {\bibfnamefont {P.}~\bibnamefont
  {Ovartchaiyapong}}, \bibinfo {author} {\bibfnamefont {L.~M.~A.}\ \bibnamefont
  {Pascal}}, \bibinfo {author} {\bibfnamefont {B.~A.}\ \bibnamefont {Myers}},
  \bibinfo {author} {\bibfnamefont {P.}~\bibnamefont {Lauria}}, \ and\ \bibinfo
  {author} {\bibfnamefont {A.~C.~Bleszynski}\ \bibnamefont {Jayich}},\
  }\bibfield  {title} {\enquote {\bibinfo {title} {High quality factor
  single-crystal diamond mechanical resonators},}\ }\href@noop {} {\bibfield
  {journal} {\bibinfo  {journal} {Appl. Phys. Lett.}\ }\textbf {\bibinfo
  {volume} {101}},\ \bibinfo {pages} {163505} (\bibinfo {year}
  {2012})}\BibitemShut {NoStop}%
\bibitem [{\citenamefont {Bunch}\ \emph {et~al.}(2007)\citenamefont {Bunch},
  \citenamefont {van~der Zande}, \citenamefont {Verbridge}, \citenamefont
  {Frank}, \citenamefont {Tanenbaum}, \citenamefont {Parpia}, \citenamefont
  {Craighead},\ and\ \citenamefont {McEuen}}]{SCI-315-490}%
  \BibitemOpen
  \bibfield  {author} {\bibinfo {author} {\bibfnamefont {J.~Scott}\
  \bibnamefont {Bunch}}, \bibinfo {author} {\bibfnamefont {Arend~M.}\
  \bibnamefont {van~der Zande}}, \bibinfo {author} {\bibfnamefont {Scott~S.}\
  \bibnamefont {Verbridge}}, \bibinfo {author} {\bibfnamefont {Ian~W.}\
  \bibnamefont {Frank}}, \bibinfo {author} {\bibfnamefont {David~M.}\
  \bibnamefont {Tanenbaum}}, \bibinfo {author} {\bibfnamefont {Jeevak~M.}\
  \bibnamefont {Parpia}}, \bibinfo {author} {\bibfnamefont {Harold~G.}\
  \bibnamefont {Craighead}}, \ and\ \bibinfo {author} {\bibfnamefont {Paul~L.}\
  \bibnamefont {McEuen}},\ }\bibfield  {title} {\enquote {\bibinfo {title}
  {Electromechanical resonators from graphene sheets},}\ }\href@noop {}
  {\bibfield  {journal} {\bibinfo  {journal} {Science}\ }\textbf {\bibinfo
  {volume} {315}},\ \bibinfo {pages} {490} (\bibinfo {year}
  {2007})}\BibitemShut {NoStop}%
\bibitem [{\citenamefont {Ganzhorn}\ \emph {et~al.}(2013)\citenamefont
  {Ganzhorn}, \citenamefont {Klyatskaya}, \citenamefont {Ruben},\ and\
  \citenamefont {Wernsdorfer}}]{Nat-Nano-8-165}%
  \BibitemOpen
  \bibfield  {author} {\bibinfo {author} {\bibfnamefont {Marc}\ \bibnamefont
  {Ganzhorn}}, \bibinfo {author} {\bibfnamefont {Svetlana}\ \bibnamefont
  {Klyatskaya}}, \bibinfo {author} {\bibfnamefont {Mario}\ \bibnamefont
  {Ruben}}, \ and\ \bibinfo {author} {\bibfnamefont {Wolfgang}\ \bibnamefont
  {Wernsdorfer}},\ }\bibfield  {title} {\enquote {\bibinfo {title} {Strong
  spin-phonon coupling between a single-molecule magnet and a carbon nanotube
  nanoelectromechanical system},}\ }\href@noop {} {\bibfield  {journal}
  {\bibinfo  {journal} {Nat. Nanotech.}\ }\textbf {\bibinfo {volume} {8}},\
  \bibinfo {pages} {165} (\bibinfo {year} {2013})}\BibitemShut {NoStop}%
\bibitem [{\citenamefont {P\'{a}lyi}\ \emph {et~al.}(2012)\citenamefont
  {P\'{a}lyi}, \citenamefont {Struck}, \citenamefont {Rudner}, \citenamefont
  {Flensberg},\ and\ \citenamefont {Burkard}}]{prl-108-206811}%
  \BibitemOpen
  \bibfield  {author} {\bibinfo {author} {\bibfnamefont {Andr\'{a}s}\
  \bibnamefont {P\'{a}lyi}}, \bibinfo {author} {\bibfnamefont {P.~R.}\
  \bibnamefont {Struck}}, \bibinfo {author} {\bibfnamefont {Mark}\ \bibnamefont
  {Rudner}}, \bibinfo {author} {\bibfnamefont {Karsten}\ \bibnamefont
  {Flensberg}}, \ and\ \bibinfo {author} {\bibfnamefont {Guido}\ \bibnamefont
  {Burkard}},\ }\bibfield  {title} {\enquote {\bibinfo {title}
  {Spin-orbit-induced strong coupling of a single spin to a nanomechanical
  resonator},}\ }\href@noop {} {\bibfield  {journal} {\bibinfo  {journal}
  {Phys.\ Rev. Lett.}\ }\textbf {\bibinfo {volume} {108}},\ \bibinfo {pages}
  {206811} (\bibinfo {year} {2012})}\BibitemShut {NoStop}%
\bibitem [{\citenamefont {Eichler}\ \emph {et~al.}(2012)\citenamefont
  {Eichler}, \citenamefont {del \'{A}~lamo Ruiz}, \citenamefont {Plaza},\ and\
  \citenamefont {Bachtold}}]{prl-109-025503}%
  \BibitemOpen
  \bibfield  {author} {\bibinfo {author} {\bibfnamefont {A.}~\bibnamefont
  {Eichler}}, \bibinfo {author} {\bibfnamefont {M.}~\bibnamefont {del
  \'{A}lamo Ruiz}}, \bibinfo {author} {\bibfnamefont {J.~A.}\ \bibnamefont
  {Plaza}}, \ and\ \bibinfo {author} {\bibfnamefont {A.}~\bibnamefont
  {Bachtold}},\ }\bibfield  {title} {\enquote {\bibinfo {title} {Strong
  coupling between mechanical modes in a nanotube resonator},}\ }\href@noop {}
  {\bibfield  {journal} {\bibinfo  {journal} {Phys.\ Rev. Lett.}\ }\textbf
  {\bibinfo {volume} {109}},\ \bibinfo {pages} {025503} (\bibinfo {year}
  {2012})}\BibitemShut {NoStop}%
\bibitem [{\citenamefont {K\'{a}lm\'{a}n}\ \emph {et~al.}(2012)\citenamefont
  {K\'{a}lm\'{a}n}, \citenamefont {Kiss}, \citenamefont {Fort\'{a}gh},\ and\
  \citenamefont {Domokos}}]{nl-12-435}%
  \BibitemOpen
  \bibfield  {author} {\bibinfo {author} {\bibfnamefont {O.}~\bibnamefont
  {K\'{a}lm\'{a}n}}, \bibinfo {author} {\bibfnamefont {T.}~\bibnamefont
  {Kiss}}, \bibinfo {author} {\bibfnamefont {J.}~\bibnamefont {Fort\'{a}gh}}, \
  and\ \bibinfo {author} {\bibfnamefont {P.}~\bibnamefont {Domokos}},\
  }\bibfield  {title} {\enquote {\bibinfo {title} {Quantum galvanometer by
  interfacing a vibrating nanowire and cold atoms},}\ }\href@noop {} {\bibfield
   {journal} {\bibinfo  {journal} {Nano Lett.}\ }\textbf {\bibinfo {volume}
  {12}},\ \bibinfo {pages} {435} (\bibinfo {year} {2012})}\BibitemShut
  {NoStop}%
\bibitem [{\citenamefont {Dar\'{a}zs}\ \emph {et~al.}(2014)\citenamefont
  {Dar\'{a}zs}, \citenamefont {Kurucz}, \citenamefont {K\'{a}lm\'{a}n},
  \citenamefont {Kiss}, \citenamefont {Fort\'{a}gh},\ and\ \citenamefont
  {Domokos}}]{prl-112-133603}%
  \BibitemOpen
  \bibfield  {author} {\bibinfo {author} {\bibfnamefont {Z.}~\bibnamefont
  {Dar\'{a}zs}}, \bibinfo {author} {\bibfnamefont {Z.}~\bibnamefont {Kurucz}},
  \bibinfo {author} {\bibfnamefont {O.}~\bibnamefont {K\'{a}lm\'{a}n}},
  \bibinfo {author} {\bibfnamefont {T.}~\bibnamefont {Kiss}}, \bibinfo {author}
  {\bibfnamefont {J.}~\bibnamefont {Fort\'{a}gh}}, \ and\ \bibinfo {author}
  {\bibfnamefont {P.}~\bibnamefont {Domokos}},\ }\bibfield  {title} {\enquote
  {\bibinfo {title} {Parametric amplification of the mechanical vibrations of a
  suspended nanowire by magnetic coupling to a \text{Bose-Einstein}
  condensate},}\ }\href@noop {} {\bibfield  {journal} {\bibinfo  {journal}
  {Phys.\ Rev. Lett.}\ }\textbf {\bibinfo {volume} {112}},\ \bibinfo {pages}
  {133603} (\bibinfo {year} {2014})}\BibitemShut {NoStop}%
\bibitem [{\citenamefont {Aykol}\ \emph {et~al.}(2014)\citenamefont {Aykol},
  \citenamefont {Hou}, \citenamefont {Dhall}, \citenamefont {Chang},
  \citenamefont {Branham}, \citenamefont {Qiu},\ and\ \citenamefont
  {Cronin}}]{nl-14-2426}%
  \BibitemOpen
  \bibfield  {author} {\bibinfo {author} {\bibfnamefont {Mehmet}\ \bibnamefont
  {Aykol}}, \bibinfo {author} {\bibfnamefont {Bingya}\ \bibnamefont {Hou}},
  \bibinfo {author} {\bibfnamefont {Rohan}\ \bibnamefont {Dhall}}, \bibinfo
  {author} {\bibfnamefont {Shun-Wen}\ \bibnamefont {Chang}}, \bibinfo {author}
  {\bibfnamefont {William}\ \bibnamefont {Branham}}, \bibinfo {author}
  {\bibfnamefont {Jing}\ \bibnamefont {Qiu}}, \ and\ \bibinfo {author}
  {\bibfnamefont {Stephen~B.}\ \bibnamefont {Cronin}},\ }\bibfield  {title}
  {\enquote {\bibinfo {title} {Clamping instability and van der \text{W}aals
  forces in carbon nanotube mechanical resonators},}\ }\href@noop {} {\bibfield
   {journal} {\bibinfo  {journal} {Nano Lett.}\ }\textbf {\bibinfo {volume}
  {14}},\ \bibinfo {pages} {2426} (\bibinfo {year} {2014})}\BibitemShut
  {NoStop}%
\bibitem [{\citenamefont {Stadler}\ \emph {et~al.}(2014)\citenamefont
  {Stadler}, \citenamefont {Belzig},\ and\ \citenamefont
  {Rastelli}}]{prl-113-047201}%
  \BibitemOpen
  \bibfield  {author} {\bibinfo {author} {\bibfnamefont {P.}~\bibnamefont
  {Stadler}}, \bibinfo {author} {\bibfnamefont {W.}~\bibnamefont {Belzig}}, \
  and\ \bibinfo {author} {\bibfnamefont {G.}~\bibnamefont {Rastelli}},\
  }\bibfield  {title} {\enquote {\bibinfo {title} {Ground-state cooling of a
  carbon nanomechanical resonator by spin-polarized current},}\ }\href@noop {}
  {\bibfield  {journal} {\bibinfo  {journal} {Phys.\ Rev. Lett.}\ }\textbf
  {\bibinfo {volume} {113}},\ \bibinfo {pages} {047201} (\bibinfo {year}
  {2014})}\BibitemShut {NoStop}%
\bibitem [{\citenamefont {Singh}\ \emph {et~al.}(2014)\citenamefont {Singh},
  \citenamefont {Bosman}, \citenamefont {Schneider}, \citenamefont {Blanter},
  \citenamefont {Castellanos-Gomez},\ and\ \citenamefont
  {Steele}}]{Nat-Nano-9-820}%
  \BibitemOpen
  \bibfield  {author} {\bibinfo {author} {\bibfnamefont {V.}~\bibnamefont
  {Singh}}, \bibinfo {author} {\bibfnamefont {S.~J.}\ \bibnamefont {Bosman}},
  \bibinfo {author} {\bibfnamefont {B.~H.}\ \bibnamefont {Schneider}}, \bibinfo
  {author} {\bibfnamefont {Y.~M.}\ \bibnamefont {Blanter}}, \bibinfo {author}
  {\bibfnamefont {A.}~\bibnamefont {Castellanos-Gomez}}, \ and\ \bibinfo
  {author} {\bibfnamefont {G.~A.}\ \bibnamefont {Steele}},\ }\bibfield  {title}
  {\enquote {\bibinfo {title} {Optomechanical coupling between a multilayer
  graphene mechanical resonator and a superconducting microwave cavity},}\
  }\href@noop {} {\bibfield  {journal} {\bibinfo  {journal} {Nat. Nanotech.}\
  }\textbf {\bibinfo {volume} {9}},\ \bibinfo {pages} {820} (\bibinfo {year}
  {2014})}\BibitemShut {NoStop}%
\bibitem [{\citenamefont {Weber}\ \emph {et~al.}(2014)\citenamefont {Weber},
  \citenamefont {Guttinger}, \citenamefont {Tsioutsios}, \citenamefont
  {Chang},\ and\ \citenamefont {Bachtold}}]{nl-14-2854}%
  \BibitemOpen
  \bibfield  {author} {\bibinfo {author} {\bibfnamefont {P.}~\bibnamefont
  {Weber}}, \bibinfo {author} {\bibfnamefont {J.}~\bibnamefont {G\"{u}ttinger}},
  \bibinfo {author} {\bibfnamefont {I.}~\bibnamefont {Tsioutsios}}, \bibinfo
  {author} {\bibfnamefont {D.~E.}\ \bibnamefont {Chang}}, \ and\ \bibinfo
  {author} {\bibfnamefont {A.}~\bibnamefont {Bachtold}},\ }\bibfield  {title}
  {\enquote {\bibinfo {title} {Coupling graphene mechanical resonators to
  superconducting microwave cavities},}\ }\href@noop {} {\bibfield  {journal}
  {\bibinfo  {journal} {Nano Lett.}\ }\textbf {\bibinfo {volume} {14}},\
  \bibinfo {pages} {2854} (\bibinfo {year} {2014})}\BibitemShut {NoStop}%
\bibitem [{\citenamefont {Muschik}\ \emph {et~al.}(2014)\citenamefont
  {Muschik}, \citenamefont {Moulieras}, \citenamefont {Bachtold}, \citenamefont
  {Koppens}, \citenamefont {Lewenstein},\ and\ \citenamefont
  {Chang}}]{prl-112-223601}%
  \BibitemOpen
  \bibfield  {author} {\bibinfo {author} {\bibfnamefont {Christine~A.}\
  \bibnamefont {Muschik}}, \bibinfo {author} {\bibfnamefont {Simon}\
  \bibnamefont {Moulieras}}, \bibinfo {author} {\bibfnamefont {Adrian}\
  \bibnamefont {Bachtold}}, \bibinfo {author} {\bibfnamefont {Frank H.~L.}\
  \bibnamefont {Koppens}}, \bibinfo {author} {\bibfnamefont {Maciej}\
  \bibnamefont {Lewenstein}}, \ and\ \bibinfo {author} {\bibfnamefont
  {Darrick~E.}\ \bibnamefont {Chang}},\ }\bibfield  {title} {\enquote {\bibinfo
  {title} {Harnessing vacuum forces for quantum sensing of graphene motion},}\
  }\href@noop {} {\bibfield  {journal} {\bibinfo  {journal} {Phys.\ Rev.
  Lett.}\ }\textbf {\bibinfo {volume} {112}},\ \bibinfo {pages} {223601}
  (\bibinfo {year} {2014})}\BibitemShut {NoStop}%
\bibitem [{\citenamefont {Reserbat-Plantey}\ \emph {et~al.}()\citenamefont
  {Reserbat-Plantey}, \citenamefont {Schadler}, \citenamefont {Gaudreau},
  \citenamefont {Navickaite}, \citenamefont {Guttinger}, \citenamefont {Chang},
  \citenamefont {Toninelli}, \citenamefont {Bachtold},\ and\ \citenamefont
  {Koppens}}]{eprint-1504-08275}%
  \BibitemOpen
  \bibfield  {author} {\bibinfo {author} {\bibfnamefont {Antoine}\ \bibnamefont
  {Reserbat-Plantey}}, \bibinfo {author} {\bibfnamefont {Kevin~G.}\
  \bibnamefont {Schadler}}, \bibinfo {author} {\bibfnamefont {Louis}\
  \bibnamefont {Gaudreau}}, \bibinfo {author} {\bibfnamefont {Gabriele}\
  \bibnamefont {Navickaite}}, \bibinfo {author} {\bibfnamefont {Johannes}\
  \bibnamefont {Guttinger}}, \bibinfo {author} {\bibfnamefont {Darrick}\
  \bibnamefont {Chang}}, \bibinfo {author} {\bibfnamefont {Costanza}\
  \bibnamefont {Toninelli}}, \bibinfo {author} {\bibfnamefont {Adrian}\
  \bibnamefont {Bachtold}}, \ and\ \bibinfo {author} {\bibfnamefont
  {Frank~H.L.}\ \bibnamefont {Koppens}},\ }\href@noop {} {\enquote {\bibinfo
  {title} {Electro-mechanical control of an optical emitter using graphene},}\
  }\Eprint {http://arxiv.org/abs/arXiv:1504.08275 [cond-mat.mes-hall]}
  {arXiv:1504.08275 [cond-mat.mes-hall]} \BibitemShut {NoStop}%
\bibitem [{\citenamefont {Rabl}\ \emph {et~al.}(2009)\citenamefont {Rabl},
  \citenamefont {Cappellaro}, \citenamefont {Dutt}, \citenamefont {Jiang},
  \citenamefont {Maze},\ and\ \citenamefont {Lukin}}]{prb-79-041302}%
  \BibitemOpen
  \bibfield  {author} {\bibinfo {author} {\bibfnamefont {P.}~\bibnamefont
  {Rabl}}, \bibinfo {author} {\bibfnamefont {P.}~\bibnamefont {Cappellaro}},
  \bibinfo {author} {\bibfnamefont {M.~V.~Gurudev}\ \bibnamefont {Dutt}},
  \bibinfo {author} {\bibfnamefont {L.}~\bibnamefont {Jiang}}, \bibinfo
  {author} {\bibfnamefont {J.~R.}\ \bibnamefont {Maze}}, \ and\ \bibinfo
  {author} {\bibfnamefont {M.~D.}\ \bibnamefont {Lukin}},\ }\bibfield  {title}
  {\enquote {\bibinfo {title} {Strong magnetic coupling between an electronic
  spin qubit and a mechanical resonator},}\ }\href@noop {} {\bibfield
  {journal} {\bibinfo  {journal} {Phys.\ Rev. B}\ }\textbf {\bibinfo {volume}
  {79}},\ \bibinfo {pages} {041302(R)} (\bibinfo {year} {2009})}\BibitemShut
  {NoStop}%
\bibitem [{\citenamefont {Rabl}\ \emph {et~al.}(2010)\citenamefont {Rabl},
  \citenamefont {Kolkowitz}, \citenamefont {Koppens}, \citenamefont {Harris},
  \citenamefont {Zoller},\ and\ \citenamefont {Lukin}}]{natphys-6-602}%
  \BibitemOpen
  \bibfield  {author} {\bibinfo {author} {\bibfnamefont {P.}~\bibnamefont
  {Rabl}}, \bibinfo {author} {\bibfnamefont {S.~J.}\ \bibnamefont {Kolkowitz}},
  \bibinfo {author} {\bibfnamefont {F.~H.~L.}\ \bibnamefont {Koppens}},
  \bibinfo {author} {\bibfnamefont {J.~G.~E.}\ \bibnamefont {Harris}}, \bibinfo
  {author} {\bibfnamefont {P.}~\bibnamefont {Zoller}}, \ and\ \bibinfo {author}
  {\bibfnamefont {M.~D.}\ \bibnamefont {Lukin}},\ }\bibfield  {title} {\enquote
  {\bibinfo {title} {A quantum spin transducer based on nanoelectromechanical
  resonator arrays},}\ }\href@noop {} {\bibfield  {journal} {\bibinfo
  {journal} {Nat. Phys.}\ }\textbf {\bibinfo {volume} {6}},\ \bibinfo {pages}
  {602} (\bibinfo {year} {2010})}\BibitemShut {NoStop}%
\bibitem [{\citenamefont {Arcizet}\ \emph {et~al.}(2011)\citenamefont
  {Arcizet}, \citenamefont {Jacques}, \citenamefont {Siria}, \citenamefont
  {Poncharal}, \citenamefont {Vincent},\ and\ \citenamefont
  {Seidelin}}]{np-7-879}%
  \BibitemOpen
  \bibfield  {author} {\bibinfo {author} {\bibfnamefont {O.}~\bibnamefont
  {Arcizet}}, \bibinfo {author} {\bibfnamefont {V.}~\bibnamefont {Jacques}},
  \bibinfo {author} {\bibfnamefont {A.}~\bibnamefont {Siria}}, \bibinfo
  {author} {\bibfnamefont {P.}~\bibnamefont {Poncharal}}, \bibinfo {author}
  {\bibfnamefont {P.}~\bibnamefont {Vincent}}, \ and\ \bibinfo {author}
  {\bibfnamefont {S.}~\bibnamefont {Seidelin}},\ }\bibfield  {title} {\enquote
  {\bibinfo {title} {A single nitrogen-vacancy defect coupled to a
  nanomechanical oscillator},}\ }\href@noop {} {\bibfield  {journal} {\bibinfo
  {journal} {Nat. Phys.}\ }\textbf {\bibinfo {volume} {7}},\ \bibinfo {pages}
  {879} (\bibinfo {year} {2011})}\BibitemShut {NoStop}%
\bibitem [{\citenamefont {Kolkowitz}\ \emph {et~al.}(2012)\citenamefont
  {Kolkowitz}, \citenamefont {Jayich}, \citenamefont {Unterreithmeier},
  \citenamefont {Bennett}, \citenamefont {Rabl}, \citenamefont {Harris},\ and\
  \citenamefont {Lukin}}]{SCI-335-1603}%
  \BibitemOpen
  \bibfield  {author} {\bibinfo {author} {\bibfnamefont {Shimon}\ \bibnamefont
  {Kolkowitz}}, \bibinfo {author} {\bibfnamefont {Ania C.~Bleszynski}\
  \bibnamefont {Jayich}}, \bibinfo {author} {\bibfnamefont {Quirin~P.}\
  \bibnamefont {Unterreithmeier}}, \bibinfo {author} {\bibfnamefont
  {Steven~D.}\ \bibnamefont {Bennett}}, \bibinfo {author} {\bibfnamefont
  {Peter}\ \bibnamefont {Rabl}}, \bibinfo {author} {\bibfnamefont {J.~G.~E.}\
  \bibnamefont {Harris}}, \ and\ \bibinfo {author} {\bibfnamefont {Mikhail~D.}\
  \bibnamefont {Lukin}},\ }\bibfield  {title} {\enquote {\bibinfo {title}
  {Coherent sensing of a mechanical resonator with a single-spin qubit},}\
  }\href@noop {} {\bibfield  {journal} {\bibinfo  {journal} {Science}\ }\textbf
  {\bibinfo {volume} {335}},\ \bibinfo {pages} {1603} (\bibinfo {year}
  {2012})}\BibitemShut {NoStop}%
\bibitem [{\citenamefont {Hong}\ \emph {et~al.}(2012)\citenamefont {Hong},
  \citenamefont {Grinolds}, \citenamefont {Maletinsky}, \citenamefont
  {Walsworth}, \citenamefont {Lukin},\ and\ \citenamefont {Yacoby}}]{nl-12}%
  \BibitemOpen
  \bibfield  {author} {\bibinfo {author} {\bibfnamefont {S.}~\bibnamefont
  {Hong}}, \bibinfo {author} {\bibfnamefont {M.~S.}\ \bibnamefont {Grinolds}},
  \bibinfo {author} {\bibfnamefont {P.}~\bibnamefont {Maletinsky}}, \bibinfo
  {author} {\bibfnamefont {R.~L.}\ \bibnamefont {Walsworth}}, \bibinfo {author}
  {\bibfnamefont {M.~D.}\ \bibnamefont {Lukin}}, \ and\ \bibinfo {author}
  {\bibfnamefont {A.}~\bibnamefont {Yacoby}},\ }\bibfield  {title} {\enquote
  {\bibinfo {title} {Coherent, mechanical control of a single electronic
  spin},}\ }\href@noop {} {\bibfield  {journal} {\bibinfo  {journal} {Nano
  Lett.}\ }\textbf {\bibinfo {volume} {12}},\ \bibinfo {pages} {3920} (\bibinfo
  {year} {2012})}\BibitemShut {NoStop}%
\bibitem [{\citenamefont {Pigeau}\ \emph {et~al.}(2015)\citenamefont {Pigeau},
  \citenamefont {Rohr}, \citenamefont {de~Lepinay}, \citenamefont {Gloppe},
  \citenamefont {Jacques},\ and\ \citenamefont {Arcizet}}]{NC-6-8603}%
  \BibitemOpen
  \bibfield  {author} {\bibinfo {author} {\bibfnamefont {B.}~\bibnamefont
  {Pigeau}}, \bibinfo {author} {\bibfnamefont {S.}~\bibnamefont {Rohr}},
  \bibinfo {author} {\bibfnamefont {L.~Mercier}\ \bibnamefont {de~Lepinay}},
  \bibinfo {author} {\bibfnamefont {A.}~\bibnamefont {Gloppe}}, \bibinfo
  {author} {\bibfnamefont {V.}~\bibnamefont {Jacques}}, \ and\ \bibinfo
  {author} {\bibfnamefont {O.}~\bibnamefont {Arcizet}},\ }\bibfield  {title}
  {\enquote {\bibinfo {title} {Observation of a phononic mollow triplet in a
  multimode hybrid spin-nanomechanical system},}\ }\href@noop {} {\bibfield
  {journal} {\bibinfo  {journal} {Nat. Commun.}\ }\textbf {\bibinfo {volume}
  {6}},\ \bibinfo {pages} {8603} (\bibinfo {year} {2015})}\BibitemShut
  {NoStop}%
\bibitem [{\citenamefont {Li}\ \emph {et~al.}(2015)\citenamefont {Li},
  \citenamefont {Liu}, \citenamefont {Gao}, \citenamefont {Xiang},
  \citenamefont {Rabl}, \citenamefont {Xiao},\ and\ \citenamefont
  {Li}}]{prappl-4-044003}%
  \BibitemOpen
  \bibfield  {author} {\bibinfo {author} {\bibfnamefont {Peng-Bo}\ \bibnamefont
  {Li}}, \bibinfo {author} {\bibfnamefont {Yong-Chun}\ \bibnamefont {Liu}},
  \bibinfo {author} {\bibfnamefont {S.-Y.}\ \bibnamefont {Gao}}, \bibinfo
  {author} {\bibfnamefont {Ze-Liang}\ \bibnamefont {Xiang}}, \bibinfo {author}
  {\bibfnamefont {Peter}\ \bibnamefont {Rabl}}, \bibinfo {author}
  {\bibfnamefont {Yun-Feng}\ \bibnamefont {Xiao}}, \ and\ \bibinfo {author}
  {\bibfnamefont {Fu-Li}\ \bibnamefont {Li}},\ }\bibfield  {title} {\enquote
  {\bibinfo {title} {Hybrid quantum device based on \text{NV} centers in
  diamond nanomechanical resonators plus superconducting waveguide cavities},}\
  }\href@noop {} {\bibfield  {journal} {\bibinfo  {journal} {Phys.\ Rev.
  Applied}\ }\textbf {\bibinfo {volume} {4}},\ \bibinfo {pages} {044003}
  (\bibinfo {year} {2015})}\BibitemShut {NoStop}%
\bibitem [{\citenamefont {Xu}\ \emph {et~al.}(2009)\citenamefont {Xu},
  \citenamefont {Hu}, \citenamefont {Yang}, \citenamefont {Feng},\ and\
  \citenamefont {Du}}]{pra-80-022335}%
  \BibitemOpen
  \bibfield  {author} {\bibinfo {author} {\bibfnamefont {Z.~Y.}\ \bibnamefont
  {Xu}}, \bibinfo {author} {\bibfnamefont {Y.~M.}\ \bibnamefont {Hu}}, \bibinfo
  {author} {\bibfnamefont {W.~L.}\ \bibnamefont {Yang}}, \bibinfo {author}
  {\bibfnamefont {M.}~\bibnamefont {Feng}}, \ and\ \bibinfo {author}
  {\bibfnamefont {J.~F.}\ \bibnamefont {Du}},\ }\bibfield  {title} {\enquote
  {\bibinfo {title} {Deterministically entangling distant nitrogen-vacancy
  centers by a nanomechanical cantilever},}\ }\href@noop {} {\bibfield
  {journal} {\bibinfo  {journal} {Phys.\ Rev. A}\ }\textbf {\bibinfo {volume}
  {80}},\ \bibinfo {pages} {022335} (\bibinfo {year} {2009})}\BibitemShut
  {NoStop}%
\bibitem [{\citenamefont {Zhou}\ \emph {et~al.}(2010)\citenamefont {Zhou},
  \citenamefont {Wei}, \citenamefont {Gao},\ and\ \citenamefont
  {Wang}}]{pra-81-042323}%
  \BibitemOpen
  \bibfield  {author} {\bibinfo {author} {\bibfnamefont {L.~G.}\ \bibnamefont
  {Zhou}}, \bibinfo {author} {\bibfnamefont {L.~F.}\ \bibnamefont {Wei}},
  \bibinfo {author} {\bibfnamefont {M.}~\bibnamefont {Gao}}, \ and\ \bibinfo
  {author} {\bibfnamefont {X.~B.}\ \bibnamefont {Wang}},\ }\bibfield  {title}
  {\enquote {\bibinfo {title} {Strong coupling between two distant electronic
  spins via a nanomechanical resonator},}\ }\href@noop {} {\bibfield  {journal}
  {\bibinfo  {journal} {Phys.\ Rev. A}\ }\textbf {\bibinfo {volume} {81}},\
  \bibinfo {pages} {042323} (\bibinfo {year} {2010})}\BibitemShut {NoStop}%
\bibitem [{\citenamefont {Yin}\ \emph {et~al.}(2013)\citenamefont {Yin},
  \citenamefont {Li}, \citenamefont {Zhang},\ and\ \citenamefont
  {Duan}}]{pra-88-033614}%
  \BibitemOpen
  \bibfield  {author} {\bibinfo {author} {\bibfnamefont {Z.~Q.}\ \bibnamefont
  {Yin}}, \bibinfo {author} {\bibfnamefont {T.~C.}\ \bibnamefont {Li}},
  \bibinfo {author} {\bibfnamefont {X.}~\bibnamefont {Zhang}}, \ and\ \bibinfo
  {author} {\bibfnamefont {L.~M.}\ \bibnamefont {Duan}},\ }\bibfield  {title}
  {\enquote {\bibinfo {title} {Large quantum superpositions of a levitated
  nanodiamond through spin-optomechanical coupling},}\ }\href@noop {}
  {\bibfield  {journal} {\bibinfo  {journal} {Phys.\ Rev. A}\ }\textbf
  {\bibinfo {volume} {88}},\ \bibinfo {pages} {033614} (\bibinfo {year}
  {2013})}\BibitemShut {NoStop}%
\bibitem [{\citenamefont {Chotorlishvili}\ \emph {et~al.}(2013)\citenamefont
  {Chotorlishvili}, \citenamefont {Sander}, \citenamefont {Sukhov},
  \citenamefont {Dugaev}, \citenamefont {Vieira}, \citenamefont {Komnik},\ and\
  \citenamefont {Berakdar}}]{prb-88-085201}%
  \BibitemOpen
  \bibfield  {author} {\bibinfo {author} {\bibfnamefont {L.}~\bibnamefont
  {Chotorlishvili}}, \bibinfo {author} {\bibfnamefont {D.}~\bibnamefont
  {Sander}}, \bibinfo {author} {\bibfnamefont {A.}~\bibnamefont {Sukhov}},
  \bibinfo {author} {\bibfnamefont {V.}~\bibnamefont {Dugaev}}, \bibinfo
  {author} {\bibfnamefont {V.~R.}\ \bibnamefont {Vieira}}, \bibinfo {author}
  {\bibfnamefont {A.}~\bibnamefont {Komnik}}, \ and\ \bibinfo {author}
  {\bibfnamefont {J.}~\bibnamefont {Berakdar}},\ }\bibfield  {title} {\enquote
  {\bibinfo {title} {Entanglement between nitrogen vacancy spins in diamond
  controlled by a nanomechanical resonator},}\ }\href@noop {} {\bibfield
  {journal} {\bibinfo  {journal} {Phys.\ Rev. B}\ }\textbf {\bibinfo {volume}
  {88}},\ \bibinfo {pages} {085201} (\bibinfo {year} {2013})}\BibitemShut
  {NoStop}%
\bibitem [{\citenamefont {Bennett}\ \emph {et~al.}(2013)\citenamefont
  {Bennett}, \citenamefont {Yao}, \citenamefont {Otterbach}, \citenamefont
  {Zoller}, \citenamefont {Rabl},\ and\ \citenamefont
  {Lukin}}]{prl-110-156402}%
  \BibitemOpen
  \bibfield  {author} {\bibinfo {author} {\bibfnamefont {S.~D.}\ \bibnamefont
  {Bennett}}, \bibinfo {author} {\bibfnamefont {N.Y.}\ \bibnamefont {Yao}},
  \bibinfo {author} {\bibfnamefont {J.}~\bibnamefont {Otterbach}}, \bibinfo
  {author} {\bibfnamefont {P.}~\bibnamefont {Zoller}}, \bibinfo {author}
  {\bibfnamefont {P.}~\bibnamefont {Rabl}}, \ and\ \bibinfo {author}
  {\bibfnamefont {M.~D.}\ \bibnamefont {Lukin}},\ }\bibfield  {title} {\enquote
  {\bibinfo {title} {Phonon-induced spin-spin interactions in diamond
  nanostructures: Application to spin squeezing},}\ }\href@noop {} {\bibfield
  {journal} {\bibinfo  {journal} {Phys.\ Rev. Lett.}\ }\textbf {\bibinfo
  {volume} {110}},\ \bibinfo {pages} {156402} (\bibinfo {year}
  {2013})}\BibitemShut {NoStop}%
\bibitem [{\citenamefont {MacQuarrie}\ \emph {et~al.}(2013)\citenamefont
  {MacQuarrie}, \citenamefont {Gosavi}, \citenamefont {Jungwirth},
  \citenamefont {Bhave},\ and\ \citenamefont {Fuchs}}]{prl-111-227602}%
  \BibitemOpen
  \bibfield  {author} {\bibinfo {author} {\bibfnamefont {E.~R.}\ \bibnamefont
  {MacQuarrie}}, \bibinfo {author} {\bibfnamefont {T.~A.}\ \bibnamefont
  {Gosavi}}, \bibinfo {author} {\bibfnamefont {N.~R.}\ \bibnamefont
  {Jungwirth}}, \bibinfo {author} {\bibfnamefont {S.~A.}\ \bibnamefont
  {Bhave}}, \ and\ \bibinfo {author} {\bibfnamefont {G.~D.}\ \bibnamefont
  {Fuchs}},\ }\bibfield  {title} {\enquote {\bibinfo {title} {Mechanical spin
  control of nitrogen-vacancy centers in diamond},}\ }\href@noop {} {\bibfield
  {journal} {\bibinfo  {journal} {Phys.\ Rev. Lett.}\ }\textbf {\bibinfo
  {volume} {111}},\ \bibinfo {pages} {227602} (\bibinfo {year}
  {2013})}\BibitemShut {NoStop}%
\bibitem [{\citenamefont {Teissier}\ \emph {et~al.}(2014)\citenamefont
  {Teissier}, \citenamefont {Barfuss}, \citenamefont {Appel}, \citenamefont
  {Neu},\ and\ \citenamefont {Maletinsky}}]{prl-113-020503}%
  \BibitemOpen
  \bibfield  {author} {\bibinfo {author} {\bibfnamefont {J.}~\bibnamefont
  {Teissier}}, \bibinfo {author} {\bibfnamefont {A.}~\bibnamefont {Barfuss}},
  \bibinfo {author} {\bibfnamefont {P.}~\bibnamefont {Appel}}, \bibinfo
  {author} {\bibfnamefont {E.}~\bibnamefont {Neu}}, \ and\ \bibinfo {author}
  {\bibfnamefont {P.}~\bibnamefont {Maletinsky}},\ }\bibfield  {title}
  {\enquote {\bibinfo {title} {Strain coupling of a nitrogen-vacancy center
  spin to a diamond mechanical oscillator},}\ }\href@noop {} {\bibfield
  {journal} {\bibinfo  {journal} {Phys.\ Rev. Lett.}\ }\textbf {\bibinfo
  {volume} {113}},\ \bibinfo {pages} {020503} (\bibinfo {year}
  {2014})}\BibitemShut {NoStop}%
\bibitem [{\citenamefont {Ovartchaiyapong}\ \emph {et~al.}(2014)\citenamefont
  {Ovartchaiyapong}, \citenamefont {Lee}, \citenamefont {Myers},\ and\
  \citenamefont {Jayich}}]{Natcomm-5-4429}%
  \BibitemOpen
  \bibfield  {author} {\bibinfo {author} {\bibfnamefont {Preeti}\ \bibnamefont
  {Ovartchaiyapong}}, \bibinfo {author} {\bibfnamefont {Kenneth~W.}\
  \bibnamefont {Lee}}, \bibinfo {author} {\bibfnamefont {Bryan~A.}\
  \bibnamefont {Myers}}, \ and\ \bibinfo {author} {\bibfnamefont {Ania
  C.~Bleszynski}\ \bibnamefont {Jayich}},\ }\bibfield  {title} {\enquote
  {\bibinfo {title} {Dynamic strain-mediated coupling of a single diamond spin
  to a mechanical resonator},}\ }\href@noop {} {\bibfield  {journal} {\bibinfo
  {journal} {Nat. Commun.}\ }\textbf {\bibinfo {volume} {5}},\ \bibinfo {pages}
  {4429} (\bibinfo {year} {2014})}\BibitemShut {NoStop}%
\bibitem [{\citenamefont {Kepesidis}\ \emph {et~al.}(2013)\citenamefont
  {Kepesidis}, \citenamefont {Bennett}, \citenamefont {Portolan}, \citenamefont
  {Lukin},\ and\ \citenamefont {Rabl}}]{prb-88-064105}%
  \BibitemOpen
  \bibfield  {author} {\bibinfo {author} {\bibfnamefont {K.~V.}\ \bibnamefont
  {Kepesidis}}, \bibinfo {author} {\bibfnamefont {S.~D.}\ \bibnamefont
  {Bennett}}, \bibinfo {author} {\bibfnamefont {S.}~\bibnamefont {Portolan}},
  \bibinfo {author} {\bibfnamefont {M.~D.}\ \bibnamefont {Lukin}}, \ and\
  \bibinfo {author} {\bibfnamefont {P.}~\bibnamefont {Rabl}},\ }\bibfield
  {title} {\enquote {\bibinfo {title} {Phonon cooling and lasing with
  nitrogen-vacancy centers in diamond},}\ }\href@noop {} {\bibfield  {journal}
  {\bibinfo  {journal} {Phys.\ Rev. B}\ }\textbf {\bibinfo {volume} {88}},\
  \bibinfo {pages} {064105} (\bibinfo {year} {2013})}\BibitemShut {NoStop}%
\bibitem [{\citenamefont {Treutlein}\ \emph {et~al.}(2007)\citenamefont
  {Treutlein}, \citenamefont {Hunger}, \citenamefont {Camerer}, \citenamefont
  {Haansch},\ and\ \citenamefont {Reichel}}]{prl-99-140403}%
  \BibitemOpen
  \bibfield  {author} {\bibinfo {author} {\bibfnamefont {Philipp}\ \bibnamefont
  {Treutlein}}, \bibinfo {author} {\bibfnamefont {David}\ \bibnamefont
  {Hunger}}, \bibinfo {author} {\bibfnamefont {Stephan}\ \bibnamefont
  {Camerer}}, \bibinfo {author} {\bibfnamefont {Theodor~W.}\ \bibnamefont
  {H\"{a}nsch}}, \ and\ \bibinfo {author} {\bibfnamefont {Jakob}\ \bibnamefont
  {Reichel}},\ }\bibfield  {title} {\enquote {\bibinfo {title} {Bose-einstein
  condensate coupled to a nanomechanical resonator on an atom chip},}\
  }\href@noop {} {\bibfield  {journal} {\bibinfo  {journal} {Phys.\ Rev.
  Lett.}\ }\textbf {\bibinfo {volume} {99}},\ \bibinfo {pages} {140403}
  (\bibinfo {year} {2007})}\BibitemShut {NoStop}%
\bibitem [{\citenamefont {Xiang}\ \emph
  {et~al.}(2013{\natexlab{a}})\citenamefont {Xiang}, \citenamefont {Ashhab},
  \citenamefont {You},\ and\ \citenamefont {Nori}}]{RMP-1}%
  \BibitemOpen
  \bibfield  {author} {\bibinfo {author} {\bibfnamefont {Ze-Liang}\
  \bibnamefont {Xiang}}, \bibinfo {author} {\bibfnamefont {Sahel}\ \bibnamefont
  {Ashhab}}, \bibinfo {author} {\bibfnamefont {J.~Q.}\ \bibnamefont {You}}, \
  and\ \bibinfo {author} {\bibfnamefont {Franco}\ \bibnamefont {Nori}},\
  }\bibfield  {title} {\enquote {\bibinfo {title} {Hybrid quantum circuits:
  Superconducting circuits interacting with other quantum systems},}\
  }\href@noop {} {\bibfield  {journal} {\bibinfo  {journal} {Rev. Mod. Phys.}\
  }\textbf {\bibinfo {volume} {85}},\ \bibinfo {pages} {623} (\bibinfo {year}
  {2013}{\natexlab{a}})}\BibitemShut {NoStop}%
\bibitem [{\citenamefont {Zu}\ \emph {et~al.}(2014)\citenamefont {Zu},
  \citenamefont {Wang}, \citenamefont {He}, \citenamefont {Zhang},
  \citenamefont {Dai}, \citenamefont {Wang},\ and\ \citenamefont
  {Duan}}]{nature-514-72}%
  \BibitemOpen
  \bibfield  {author} {\bibinfo {author} {\bibfnamefont {C.}~\bibnamefont
  {Zu}}, \bibinfo {author} {\bibfnamefont {W.-B.}\ \bibnamefont {Wang}},
  \bibinfo {author} {\bibfnamefont {L.}~\bibnamefont {He}}, \bibinfo {author}
  {\bibfnamefont {W.-G.}\ \bibnamefont {Zhang}}, \bibinfo {author}
  {\bibfnamefont {C.-Y.}\ \bibnamefont {Dai}}, \bibinfo {author} {\bibfnamefont
  {F.}~\bibnamefont {Wang}}, \ and\ \bibinfo {author} {\bibfnamefont {L.-M.}\
  \bibnamefont {Duan}},\ }\bibfield  {title} {\enquote {\bibinfo {title}
  {Experimental realization of universal geometric quantum gates with
  solid-state spins},}\ }\href@noop {} {\bibfield  {journal} {\bibinfo
  {journal} {Nature}\ }\textbf {\bibinfo {volume} {514}},\ \bibinfo {pages}
  {72} (\bibinfo {year} {2014})}\BibitemShut {NoStop}%
\bibitem [{\citenamefont {Nemoto}\ \emph {et~al.}(2014)\citenamefont {Nemoto},
  \citenamefont {Trupke}, \citenamefont {Devitt}, \citenamefont {Stephens},
  \citenamefont {Scharfenberger}, \citenamefont {Buczak}, \citenamefont
  {Nobauer}, \citenamefont {Everitt}, \citenamefont {Schmiedmayer},\ and\
  \citenamefont {Munro}}]{prx-4-031022}%
  \BibitemOpen
  \bibfield  {author} {\bibinfo {author} {\bibfnamefont {Kae}\ \bibnamefont
  {Nemoto}}, \bibinfo {author} {\bibfnamefont {Michael}\ \bibnamefont
  {Trupke}}, \bibinfo {author} {\bibfnamefont {Simon~J.}\ \bibnamefont
  {Devitt}}, \bibinfo {author} {\bibfnamefont {Ashley~M.}\ \bibnamefont
  {Stephens}}, \bibinfo {author} {\bibfnamefont {Burkhard}\ \bibnamefont
  {Scharfenberger}}, \bibinfo {author} {\bibfnamefont {Kathrin}\ \bibnamefont
  {Buczak}}, \bibinfo {author} {\bibfnamefont {Tobias}\ \bibnamefont
  {Nobauer}}, \bibinfo {author} {\bibfnamefont {Mark~S.}\ \bibnamefont
  {Everitt}}, \bibinfo {author} {\bibfnamefont {Jorg}\ \bibnamefont
  {Schmiedmayer}}, \ and\ \bibinfo {author} {\bibfnamefont {William~J.}\
  \bibnamefont {Munro}},\ }\bibfield  {title} {\enquote {\bibinfo {title}
  {Photonic architecture for scalable quantum information processing in
  diamond},}\ }\href@noop {} {\bibfield  {journal} {\bibinfo  {journal} {Phys.\
  Rev. X}\ }\textbf {\bibinfo {volume} {4}},\ \bibinfo {pages} {031022}
  (\bibinfo {year} {2014})}\BibitemShut {NoStop}%
\bibitem [{\citenamefont {Maze}\ \emph {et~al.}(2008)\citenamefont {Maze},
  \citenamefont {Stanwix}, \citenamefont {Hodges}, \citenamefont {Hong},
  \citenamefont {Taylor}, \citenamefont {Cappellaro}, \citenamefont {Jiang},
  \citenamefont {Dutt}, \citenamefont {Togan}, \citenamefont {Zibrov},
  \citenamefont {Yacoby}, \citenamefont {Walsworth},\ and\ \citenamefont
  {Lukin}}]{nature-455-644}%
  \BibitemOpen
  \bibfield  {author} {\bibinfo {author} {\bibfnamefont {J.~R.}\ \bibnamefont
  {Maze}}, \bibinfo {author} {\bibfnamefont {P.~L.}\ \bibnamefont {Stanwix}},
  \bibinfo {author} {\bibfnamefont {J.~S.}\ \bibnamefont {Hodges}}, \bibinfo
  {author} {\bibfnamefont {S.}~\bibnamefont {Hong}}, \bibinfo {author}
  {\bibfnamefont {J.~M.}\ \bibnamefont {Taylor}}, \bibinfo {author}
  {\bibfnamefont {P.}~\bibnamefont {Cappellaro}}, \bibinfo {author}
  {\bibfnamefont {L.}~\bibnamefont {Jiang}}, \bibinfo {author} {\bibfnamefont
  {M.~V.~Gurudev}\ \bibnamefont {Dutt}}, \bibinfo {author} {\bibfnamefont
  {E.}~\bibnamefont {Togan}}, \bibinfo {author} {\bibfnamefont {A.~S.}\
  \bibnamefont {Zibrov}}, \bibinfo {author} {\bibfnamefont {A.}~\bibnamefont
  {Yacoby}}, \bibinfo {author} {\bibfnamefont {R.~L.}\ \bibnamefont
  {Walsworth}}, \ and\ \bibinfo {author} {\bibfnamefont {M.~D.}\ \bibnamefont
  {Lukin}},\ }\bibfield  {title} {\enquote {\bibinfo {title} {Nanoscale
  magnetic sensing with an individual electronic spin in diamond},}\
  }\href@noop {} {\bibfield  {journal} {\bibinfo  {journal} {Nature}\ }\textbf
  {\bibinfo {volume} {455}},\ \bibinfo {pages} {644} (\bibinfo {year}
  {2008})}\BibitemShut {NoStop}%
\bibitem [{\citenamefont {Balasubramanian}\ \emph {et~al.}(2008)\citenamefont
  {Balasubramanian}, \citenamefont {Chan}, \citenamefont {Kolesov},
  \citenamefont {Al-Hmoud}, \citenamefont {Tisler}, \citenamefont {Shin},
  \citenamefont {Kim}, \citenamefont {Wojcik}, \citenamefont {Hemmer},
  \citenamefont {Krueger}, \citenamefont {Hanke}, \citenamefont
  {Leitenstorfer}, \citenamefont {Bratschitsch}, \citenamefont {Jelezko},\ and\
  \citenamefont {Wrachtrup}}]{nature-455-648}%
  \BibitemOpen
  \bibfield  {author} {\bibinfo {author} {\bibfnamefont {G.}~\bibnamefont
  {Balasubramanian}}, \bibinfo {author} {\bibfnamefont {I.~Y.}\ \bibnamefont
  {Chan}}, \bibinfo {author} {\bibfnamefont {R.}~\bibnamefont {Kolesov}},
  \bibinfo {author} {\bibfnamefont {M.}~\bibnamefont {Al-Hmoud}}, \bibinfo
  {author} {\bibfnamefont {J.}~\bibnamefont {Tisler}}, \bibinfo {author}
  {\bibfnamefont {C.}~\bibnamefont {Shin}}, \bibinfo {author} {\bibfnamefont
  {C.}~\bibnamefont {Kim}}, \bibinfo {author} {\bibfnamefont {A.}~\bibnamefont
  {Wojcik}}, \bibinfo {author} {\bibfnamefont {P.~R.}\ \bibnamefont {Hemmer}},
  \bibinfo {author} {\bibfnamefont {A.}~\bibnamefont {Krueger}}, \bibinfo
  {author} {\bibfnamefont {T.}~\bibnamefont {Hanke}}, \bibinfo {author}
  {\bibfnamefont {A.}~\bibnamefont {Leitenstorfer}}, \bibinfo {author}
  {\bibfnamefont {R.}~\bibnamefont {Bratschitsch}}, \bibinfo {author}
  {\bibfnamefont {F.}~\bibnamefont {Jelezko}}, \ and\ \bibinfo {author}
  {\bibfnamefont {J.}~\bibnamefont {Wrachtrup}},\ }\bibfield  {title} {\enquote
  {\bibinfo {title} {Nanoscale imaging magnetometry with diamond spins under
  ambient conditions},}\ }\href@noop {} {\bibfield  {journal} {\bibinfo
  {journal} {Nature}\ }\textbf {\bibinfo {volume} {455}},\ \bibinfo {pages}
  {648} (\bibinfo {year} {2008})}\BibitemShut {NoStop}%
\bibitem [{\citenamefont {Taylor}\ \emph {et~al.}(2008)\citenamefont {Taylor},
  \citenamefont {Cappellaro}, \citenamefont {Childress}, \citenamefont {Jiang},
  \citenamefont {Budker}, \citenamefont {Hemmer}, \citenamefont {Yacoby},
  \citenamefont {Walsworth},\ and\ \citenamefont {Lukin}}]{natphys-4-810}%
  \BibitemOpen
  \bibfield  {author} {\bibinfo {author} {\bibfnamefont {J.~M.}\ \bibnamefont
  {Taylor}}, \bibinfo {author} {\bibfnamefont {P.}~\bibnamefont {Cappellaro}},
  \bibinfo {author} {\bibfnamefont {L.}~\bibnamefont {Childress}}, \bibinfo
  {author} {\bibfnamefont {L.}~\bibnamefont {Jiang}}, \bibinfo {author}
  {\bibfnamefont {D.}~\bibnamefont {Budker}}, \bibinfo {author} {\bibfnamefont
  {P.~R.}\ \bibnamefont {Hemmer}}, \bibinfo {author} {\bibfnamefont
  {A.}~\bibnamefont {Yacoby}}, \bibinfo {author} {\bibfnamefont
  {R.}~\bibnamefont {Walsworth}}, \ and\ \bibinfo {author} {\bibfnamefont
  {M.~D.}\ \bibnamefont {Lukin}},\ }\bibfield  {title} {\enquote {\bibinfo
  {title} {High-sensitivity diamond magnetometer with nanoscale resolution},}\
  }\href@noop {} {\bibfield  {journal} {\bibinfo  {journal} {Nat. Phys.}\
  }\textbf {\bibinfo {volume} {4}},\ \bibinfo {pages} {810} (\bibinfo {year}
  {2008})}\BibitemShut {NoStop}%
\bibitem [{\citenamefont {Cai}\ \emph {et~al.}(2014)\citenamefont {Cai},
  \citenamefont {Jelezko},\ and\ \citenamefont {Plenio}}]{NC-5-4065}%
  \BibitemOpen
  \bibfield  {author} {\bibinfo {author} {\bibfnamefont {Jianming}\
  \bibnamefont {Cai}}, \bibinfo {author} {\bibfnamefont {Fedor}\ \bibnamefont
  {Jelezko}}, \ and\ \bibinfo {author} {\bibfnamefont {Martin~B.}\ \bibnamefont
  {Plenio}},\ }\bibfield  {title} {\enquote {\bibinfo {title} {Hybrid sensors
  based on colour centres in diamond and piezoactive layers},}\ }\href@noop {}
  {\bibfield  {journal} {\bibinfo  {journal} {Nat. Commun.}\ }\textbf {\bibinfo
  {volume} {5}},\ \bibinfo {pages} {4065} (\bibinfo {year} {2014})}\BibitemShut
  {NoStop}%
\bibitem [{\citenamefont {Wang}\ \emph {et~al.}(2015)\citenamefont {Wang},
  \citenamefont {Yuan}, \citenamefont {Huang}, \citenamefont {Rong},
  \citenamefont {Wang}, \citenamefont {Xu}, \citenamefont {Duan}, \citenamefont
  {Ju}, \citenamefont {Shi},\ and\ \citenamefont {Du}}]{NC-6-6631}%
  \BibitemOpen
  \bibfield  {author} {\bibinfo {author} {\bibfnamefont {Pengfei}\ \bibnamefont
  {Wang}}, \bibinfo {author} {\bibfnamefont {Zhenheng}\ \bibnamefont {Yuan}},
  \bibinfo {author} {\bibfnamefont {Pu}~\bibnamefont {Huang}}, \bibinfo
  {author} {\bibfnamefont {Xing}\ \bibnamefont {Rong}}, \bibinfo {author}
  {\bibfnamefont {Mengqi}\ \bibnamefont {Wang}}, \bibinfo {author}
  {\bibfnamefont {Xiangkun}\ \bibnamefont {Xu}}, \bibinfo {author}
  {\bibfnamefont {Changkui}\ \bibnamefont {Duan}}, \bibinfo {author}
  {\bibfnamefont {Chenyong}\ \bibnamefont {Ju}}, \bibinfo {author}
  {\bibfnamefont {Fazhan}\ \bibnamefont {Shi}}, \ and\ \bibinfo {author}
  {\bibfnamefont {Jiangfeng}\ \bibnamefont {Du}},\ }\bibfield  {title}
  {\enquote {\bibinfo {title} {High-resolution vector microwave magnetometry
  based on solid-state spins in diamond},}\ }\href@noop {} {\bibfield
  {journal} {\bibinfo  {journal} {Nat. Commun.}\ }\textbf {\bibinfo {volume}
  {6}},\ \bibinfo {pages} {6631} (\bibinfo {year} {2015})}\BibitemShut
  {NoStop}%
\bibitem [{Note1()}]{Note1}%
  \BibitemOpen
  \bibinfo {note} {See Supplemental Material for more details,  which includes Refs. [50-56]}\BibitemShut
  {NoStop}%
\bibitem{book-1} L. D. Landau and E. M. Lifshitz, \emph{Theory of Elasticity} (Butterworth-Heinemann, Oxford, 1986).
\bibitem{Phys.Rep-1} Menno Poot and Herre S. J. van der Zant,``Mechanical systems in the quantum regime,"  Phys. Rep. \textbf{511}, 273 (2012).
\bibitem{prb-67-235414} S. Sapmaz, Ya. M. Blanter, L. Gurevich, and H. S. J. van der Zant, ``Carbon nanotubes as nanoelectromechanical systems,"  Phys. Rev. B \textbf{67}, 235414 (2003).
\bibitem{NJP-14 115004} S. J. M. Habraken, K. Stannigel, M. D. Lukin, P. Zoller, and P. Rabl, ``Continuous mode cooling and phonon routers
for phononic quantum networks,"  New J. Phys. \textbf{14} 115004 (2012).
\bibitem{prl-95-076803} S. Roche,  Jie Jiang,  F. Triozon,  and R. Saito, ``Quantum Dephasing in Carbon Nanotubes due to Electron-Phonon Coupling," Phys. Rev. Lett. \textbf{95}, 076803 (2005).
\bibitem{JAP-98-124301} Q. Wang, ``Wave propagation in carbon nanotubes via nonlocal continuum mechanics," J. Appl. Phys. \textbf{98}, 124301 (2005).
\bibitem{physica-E-40-2791} H. Heireche, A. Tounsi, A. Benzair,  M. Maachou, E. A. Adda Bedia, ``Sound wave propagation in single-walled carbon nanotubes
using nonlocal elasticity," Physica E \textbf{40}, 2791(2008).
\bibitem [{\citenamefont {Matsuzaki}\ \emph {et~al.}(2015)\citenamefont
  {Matsuzaki}, \citenamefont {Zhu}, \citenamefont {Kakuyanagi}, \citenamefont
  {Toida}, \citenamefont {Shimo-Oka}, \citenamefont {Mizuochi}, \citenamefont
  {Nemoto}, \citenamefont {Semba}, \citenamefont {Munro}, \citenamefont
  {Yamaguchi},\ and\ \citenamefont {Saito}}]{prl-114-120501}%
  \BibitemOpen
  \bibfield  {author} {\bibinfo {author} {\bibfnamefont {Yuichiro}\
  \bibnamefont {Matsuzaki}}, \bibinfo {author} {\bibfnamefont {Xiaobo}\
  \bibnamefont {Zhu}}, \bibinfo {author} {\bibfnamefont {Kosuke}\ \bibnamefont
  {Kakuyanagi}}, \bibinfo {author} {\bibfnamefont {Hiraku}\ \bibnamefont
  {Toida}}, \bibinfo {author} {\bibfnamefont {Takaaki}\ \bibnamefont
  {Shimo-Oka}}, \bibinfo {author} {\bibfnamefont {Norikazu}\ \bibnamefont
  {Mizuochi}}, \bibinfo {author} {\bibfnamefont {Kae}\ \bibnamefont {Nemoto}},
  \bibinfo {author} {\bibfnamefont {Kouichi}\ \bibnamefont {Semba}}, \bibinfo
  {author} {\bibfnamefont {William~J.}\ \bibnamefont {Munro}}, \bibinfo
  {author} {\bibfnamefont {Hiroshi}\ \bibnamefont {Yamaguchi}}, \ and\ \bibinfo
  {author} {\bibfnamefont {Shiro}\ \bibnamefont {Saito}},\ }\bibfield  {title}
  {\enquote {\bibinfo {title} {Improving the coherence time of a quantum system
  via a coupling to a short-lived system},}\ }\href@noop {} {\bibfield
  {journal} {\bibinfo  {journal} {Phys.\ Rev. Lett.}\ }\textbf {\bibinfo
  {volume} {114}},\ \bibinfo {pages} {120501} (\bibinfo {year}
  {2015})}\BibitemShut {NoStop}%
\bibitem [{\citenamefont {Ohashi}\ \emph {et~al.}(2013)\citenamefont {Ohashi},
  \citenamefont {Rosskopf}, \citenamefont {Watanabe}, \citenamefont {Loretz},
  \citenamefont {Tao}, \citenamefont {Hauert}, \citenamefont {Tomizawa},
  \citenamefont {Ishikawa}, \citenamefont {Ishi-Hayase}, \citenamefont
  {Shikata}, \citenamefont {Degen},\ and\ \citenamefont {Itoh}}]{nl-13-3733}%
  \BibitemOpen
  \bibfield  {author} {\bibinfo {author} {\bibfnamefont {K.}~\bibnamefont
  {Ohashi}}, \bibinfo {author} {\bibfnamefont {T.}~\bibnamefont {Rosskopf}},
  \bibinfo {author} {\bibfnamefont {H.}~\bibnamefont {Watanabe}}, \bibinfo
  {author} {\bibfnamefont {M.}~\bibnamefont {Loretz}}, \bibinfo {author}
  {\bibfnamefont {Y.}~\bibnamefont {Tao}}, \bibinfo {author} {\bibfnamefont
  {R.}~\bibnamefont {Hauert}}, \bibinfo {author} {\bibfnamefont
  {S.}~\bibnamefont {Tomizawa}}, \bibinfo {author} {\bibfnamefont
  {T.}~\bibnamefont {Ishikawa}}, \bibinfo {author} {\bibfnamefont
  {J.}~\bibnamefont {Ishi-Hayase}}, \bibinfo {author} {\bibfnamefont
  {S.}~\bibnamefont {Shikata}}, \bibinfo {author} {\bibfnamefont {C.~L.}\
  \bibnamefont {Degen}}, \ and\ \bibinfo {author} {\bibfnamefont {K.~M.}\
  \bibnamefont {Itoh}},\ }\bibfield  {title} {\enquote {\bibinfo {title}
  {Negatively charged nitrogen-vacancy centers in a 5 nm thin $^{12}\text{C}$
  diamond film},}\ }\href@noop {} {\bibfield  {journal} {\bibinfo  {journal}
  {Nano Lett.}\ }\textbf {\bibinfo {volume} {13}},\ \bibinfo {pages} {4733}
  (\bibinfo {year} {2013})}\BibitemShut {NoStop}%
\bibitem [{\citenamefont {Yao}\ \emph {et~al.}(2000)\citenamefont {Yao},
  \citenamefont {Kane},\ and\ \citenamefont {Dekker}}]{prl-84-2941}%
  \BibitemOpen
  \bibfield  {author} {\bibinfo {author} {\bibfnamefont {Zhen}\ \bibnamefont
  {Yao}}, \bibinfo {author} {\bibfnamefont {Charles~L.}\ \bibnamefont {Kane}},
  \ and\ \bibinfo {author} {\bibfnamefont {Cees}\ \bibnamefont {Dekker}},\
  }\bibfield  {title} {\enquote {\bibinfo {title} {High-field electrical
  transport in single-wall carbon nanotubes},}\ }\href@noop {} {\bibfield
  {journal} {\bibinfo  {journal} {Phys.\ Rev. Lett.}\ }\textbf {\bibinfo
  {volume} {84}},\ \bibinfo {pages} {2941} (\bibinfo {year}
  {2000})}\BibitemShut {NoStop}%
\bibitem [{\citenamefont {Collins}\ \emph {et~al.}(2001)\citenamefont
  {Collins}, \citenamefont {Hersam}, \citenamefont {Arnold}, \citenamefont
  {Martel},\ and\ \citenamefont {Avouris}}]{prl-86-3128}%
  \BibitemOpen
  \bibfield  {author} {\bibinfo {author} {\bibfnamefont {Philip~G.}\
  \bibnamefont {Collins}}, \bibinfo {author} {\bibfnamefont {M.}~\bibnamefont
  {Hersam}}, \bibinfo {author} {\bibfnamefont {M.}~\bibnamefont {Arnold}},
  \bibinfo {author} {\bibfnamefont {R.}~\bibnamefont {Martel}}, \ and\ \bibinfo
  {author} {\bibfnamefont {Ph.}\ \bibnamefont {Avouris}},\ }\bibfield  {title}
  {\enquote {\bibinfo {title} {Current saturation and electrical breakdown in
  multiwalled carbon nanotubes},}\ }\href@noop {} {\bibfield  {journal}
  {\bibinfo  {journal} {Phys.\ Rev. Lett.}\ }\textbf {\bibinfo {volume} {86}},\
  \bibinfo {pages} {3128} (\bibinfo {year} {2001})}\BibitemShut {NoStop}%
\bibitem{prl-92-106804} Ali Javey, Jing Guo, Magnus Paulsson, Qian Wang, David Mann, Mark Lundstrom, and Hongjie Dai, ``High-Field Quasiballistic Transport in Short Carbon Nanotubes,  '' Phys. Rev. Lett. \textbf{92}, 106804 (2004)
\bibitem [{\citenamefont {Dai}\ \emph {et~al.}(1996)\citenamefont {Dai},
  \citenamefont {Wong},\ and\ \citenamefont {Lieber}}]{SCI-272-523}%
  \BibitemOpen
  \bibfield  {author} {\bibinfo {author} {\bibfnamefont {Hongjie}\ \bibnamefont
  {Dai}}, \bibinfo {author} {\bibfnamefont {Eric~W.}\ \bibnamefont {Wong}}, \
  and\ \bibinfo {author} {\bibfnamefont {Charles~M.}\ \bibnamefont {Lieber}},\
  }\bibfield  {title} {\enquote {\bibinfo {title} {Probing electrical transport
  in nanomaterials: Conductivity of individual carbon nanotubes},}\ }\href@noop
  {} {\bibfield  {journal} {\bibinfo  {journal} {Science}\ }\textbf {\bibinfo
  {volume} {272}},\ \bibinfo {pages} {523} (\bibinfo {year}
  {1996})}\BibitemShut {NoStop}%
\bibitem [{\citenamefont {Frank}\ \emph {et~al.}(1998)\citenamefont {Frank},
  \citenamefont {Poncharal}, \citenamefont {Wang},\ and\ \citenamefont
  {de~Heer}}]{SCI-280-1744}%
  \BibitemOpen
  \bibfield  {author} {\bibinfo {author} {\bibfnamefont {Stefan}\ \bibnamefont
  {Frank}}, \bibinfo {author} {\bibfnamefont {Philippe}\ \bibnamefont
  {Poncharal}}, \bibinfo {author} {\bibfnamefont {Z.~L.}\ \bibnamefont {Wang}},
  \ and\ \bibinfo {author} {\bibfnamefont {Walt~A.}\ \bibnamefont {de~Heer}},\
  }\bibfield  {title} {\enquote {\bibinfo {title} {Carbon nanotube quantum
  resistors},}\ }\href@noop {} {\bibfield  {journal} {\bibinfo  {journal}
  {Science}\ }\textbf {\bibinfo {volume} {280}},\ \bibinfo {pages} {1744}
  (\bibinfo {year} {1998})}\BibitemShut {NoStop}%
\bibitem [{\citenamefont {Yu}\ \emph {et~al.}(2012)\citenamefont {Yu},
  \citenamefont {Liu}, \citenamefont {Sumant}, \citenamefont {Goyal},\ and\
  \citenamefont {Balandin}}]{nl-12-1603}%
  \BibitemOpen
  \bibfield  {author} {\bibinfo {author} {\bibfnamefont {Jie}\ \bibnamefont
  {Yu}}, \bibinfo {author} {\bibfnamefont {Guanxiong}\ \bibnamefont {Liu}},
  \bibinfo {author} {\bibfnamefont {Anirudha~V.}\ \bibnamefont {Sumant}},
  \bibinfo {author} {\bibfnamefont {Vivek}\ \bibnamefont {Goyal}}, \ and\
  \bibinfo {author} {\bibfnamefont {Alexander~A.}\ \bibnamefont {Balandin}},\
  }\bibfield  {title} {\enquote {\bibinfo {title} {Graphene-on-diamond devices
  with increased current-carrying capacity: Carbon
  s$\text{p}^2$-on-s$\text{p}^3$ technology},}\ }\href@noop {} {\bibfield
  {journal} {\bibinfo  {journal} {Nano Lett.}\ }\textbf {\bibinfo {volume}
  {12}},\ \bibinfo {pages} {1603} (\bibinfo {year} {2012})}\BibitemShut
  {NoStop}%
\bibitem [{\citenamefont {Marcos}\ \emph {et~al.}(2010)\citenamefont {Marcos},
  \citenamefont {Wubs}, \citenamefont {Taylor}, \citenamefont {Aguado},
  \citenamefont {Lukin},\ and\ \citenamefont {S{\o}rensen}}]{prl-105-210501}%
  \BibitemOpen
  \bibfield  {author} {\bibinfo {author} {\bibfnamefont {D.}~\bibnamefont
  {Marcos}}, \bibinfo {author} {\bibfnamefont {M.}~\bibnamefont {Wubs}},
  \bibinfo {author} {\bibfnamefont {J.~M.}\ \bibnamefont {Taylor}}, \bibinfo
  {author} {\bibfnamefont {R.}~\bibnamefont {Aguado}}, \bibinfo {author}
  {\bibfnamefont {M.~D.}\ \bibnamefont {Lukin}}, \ and\ \bibinfo {author}
  {\bibfnamefont {A.~S.}\ \bibnamefont {S{\o}rensen}},\ }\bibfield  {title}
  {\enquote {\bibinfo {title} {Coupling nitrogen-vacancy centers in diamond to
  superconducting flux qubits},}\ }\href@noop {} {\bibfield  {journal}
  {\bibinfo  {journal} {Phys.\ Rev. Lett.}\ }\textbf {\bibinfo {volume}
  {105}},\ \bibinfo {pages} {210501} (\bibinfo {year} {2010})}\BibitemShut
  {NoStop}%
\bibitem [{\citenamefont {Twamley}\ and\ \citenamefont
  {Barrett}(2010)}]{prb-81-033614}%
  \BibitemOpen
  \bibfield  {author} {\bibinfo {author} {\bibfnamefont {J.}~\bibnamefont
  {Twamley}}\ and\ \bibinfo {author} {\bibfnamefont {S.~D.}\ \bibnamefont
  {Barrett}},\ }\bibfield  {title} {\enquote {\bibinfo {title} {Superconducting
  cavity bus for single nitrogen-vacancy defect centers in diamond},}\
  }\href@noop {} {\bibfield  {journal} {\bibinfo  {journal} {Phys.\ Rev. B}\
  }\textbf {\bibinfo {volume} {81}},\ \bibinfo {pages} {241202(R)} (\bibinfo
  {year} {2010})}\BibitemShut {NoStop}%
\bibitem [{\citenamefont {Zhu}\ \emph {et~al.}(2011)\citenamefont {Zhu},
  \citenamefont {Saito}, \citenamefont {Kemp}, \citenamefont {Kakuyanagi},
  \citenamefont {ichi Karimoto}, \citenamefont {Nakano}, \citenamefont {Munro},
  \citenamefont {Tokura}, \citenamefont {Everitt}, \citenamefont {Nemoto},
  \citenamefont {Kasu}, \citenamefont {Mizuochi},\ and\ \citenamefont
  {Semba}}]{nature-478-221}%
  \BibitemOpen
  \bibfield  {author} {\bibinfo {author} {\bibfnamefont {Xiaobo}\ \bibnamefont
  {Zhu}}, \bibinfo {author} {\bibfnamefont {Shiro}\ \bibnamefont {Saito}},
  \bibinfo {author} {\bibfnamefont {Alexander}\ \bibnamefont {Kemp}}, \bibinfo
  {author} {\bibfnamefont {Kosuke}\ \bibnamefont {Kakuyanagi}}, \bibinfo
  {author} {\bibnamefont {Shin-ichi Karimoto}}, \bibinfo
  {author} {\bibfnamefont {Hayato}\ \bibnamefont {Nakano}}, \bibinfo {author}
  {\bibfnamefont {William~J.}\ \bibnamefont {Munro}}, \bibinfo {author}
  {\bibfnamefont {Yasuhiro}\ \bibnamefont {Tokura}}, \bibinfo {author}
  {\bibfnamefont {Mark~S.}\ \bibnamefont {Everitt}}, \bibinfo {author}
  {\bibfnamefont {Kae}\ \bibnamefont {Nemoto}}, \bibinfo {author}
  {\bibfnamefont {Makoto}\ \bibnamefont {Kasu}}, \bibinfo {author}
  {\bibfnamefont {Norikazu}\ \bibnamefont {Mizuochi}}, \ and\ \bibinfo {author}
  {\bibfnamefont {Kouichi}\ \bibnamefont {Semba}},\ }\bibfield  {title}
  {\enquote {\bibinfo {title} {Coherent coupling of a superconducting flux
  qubit to an electron spin ensemble in diamond},}\ }\href@noop {} {\bibfield
  {journal} {\bibinfo  {journal} {Nature}\ }\textbf {\bibinfo {volume} {478}},\
  \bibinfo {pages} {221} (\bibinfo {year} {2011})}\BibitemShut {NoStop}%
\bibitem [{\citenamefont {Kubo}\ \emph {et~al.}(2010)\citenamefont {Kubo},
  \citenamefont {Ong}, \citenamefont {Bertet}, \citenamefont {Vion},
  \citenamefont {Jacques}, \citenamefont {Zheng}, \citenamefont {Dreau},
  \citenamefont {Roch}, \citenamefont {Auffeves}, \citenamefont {Jelezko},
  \citenamefont {Wrachtrup}, \citenamefont {Barthe}, \citenamefont {Bergonzo},\
  and\ \citenamefont {Esteve}}]{prl-105-140502}%
  \BibitemOpen
  \bibfield  {author} {\bibinfo {author} {\bibfnamefont {Y.}~\bibnamefont
  {Kubo}}, \bibinfo {author} {\bibfnamefont {F.~R.}\ \bibnamefont {Ong}},
  \bibinfo {author} {\bibfnamefont {P.}~\bibnamefont {Bertet}}, \bibinfo
  {author} {\bibfnamefont {D.}~\bibnamefont {Vion}}, \bibinfo {author}
  {\bibfnamefont {V.}~\bibnamefont {Jacques}}, \bibinfo {author} {\bibfnamefont
  {D.}~\bibnamefont {Zheng}}, \bibinfo {author} {\bibfnamefont
  {A.}~\bibnamefont {Dreau}}, \bibinfo {author} {\bibfnamefont {J.-F.}\
  \bibnamefont {Roch}}, \bibinfo {author} {\bibfnamefont {A.}~\bibnamefont
  {Auffeves}}, \bibinfo {author} {\bibfnamefont {F.}~\bibnamefont {Jelezko}},
  \bibinfo {author} {\bibfnamefont {J.}~\bibnamefont {Wrachtrup}}, \bibinfo
  {author} {\bibfnamefont {M.~F.}\ \bibnamefont {Barthe}}, \bibinfo {author}
  {\bibfnamefont {P.}~\bibnamefont {Bergonzo}}, \ and\ \bibinfo {author}
  {\bibfnamefont {D.}~\bibnamefont {Esteve}},\ }\bibfield  {title} {\enquote
  {\bibinfo {title} {Strong coupling of a spin ensemble to a superconducting
  resonator},}\ }\href@noop {} {\bibfield  {journal} {\bibinfo  {journal}
  {Phys.\ Rev. Lett.}\ }\textbf {\bibinfo {volume} {105}},\ \bibinfo {pages}
  {140502} (\bibinfo {year} {2010})}\BibitemShut {NoStop}%
\bibitem [{\citenamefont {Xiang}\ \emph
  {et~al.}(2013{\natexlab{b}})\citenamefont {Xiang}, \citenamefont {Lu},
  \citenamefont {Li}, \citenamefont {You},\ and\ \citenamefont
  {Nori}}]{prb-87-144516}%
  \BibitemOpen
  \bibfield  {author} {\bibinfo {author} {\bibfnamefont {Ze-Liang}\
  \bibnamefont {Xiang}}, \bibinfo {author} {\bibfnamefont {Xin-You}\
  \bibnamefont {L\"{u}}}, \bibinfo {author} {\bibfnamefont {Tie-Fu}\ \bibnamefont
  {Li}}, \bibinfo {author} {\bibfnamefont {J.~Q.}\ \bibnamefont {You}}, \ and\
  \bibinfo {author} {\bibfnamefont {Franco}\ \bibnamefont {Nori}},\ }\bibfield
  {title} {\enquote {\bibinfo {title} {Hybrid quantum circuit consisting of a
  superconducting flux qubit coupled to a spin ensemble and a transmission-line
  resonator},}\ }\href@noop {} {\bibfield  {journal} {\bibinfo  {journal}
  {Phys.\ Rev. B}\ }\textbf {\bibinfo {volume} {87}},\ \bibinfo {pages}
  {144516} (\bibinfo {year} {2013}{\natexlab{b}})}\BibitemShut {NoStop}%
\bibitem [{\citenamefont {Lu}\ \emph {et~al.}(2013)\citenamefont {Lu},
  \citenamefont {Xiang}, \citenamefont {Cui}, \citenamefont {You},\ and\
  \citenamefont {Nori}}]{pra-88-012329}%
  \BibitemOpen
  \bibfield  {author} {\bibinfo {author} {\bibfnamefont {Xin-You}\ \bibnamefont
  {L\"{u}}}, \bibinfo {author} {\bibfnamefont {Ze-Liang}\ \bibnamefont {Xiang}},
  \bibinfo {author} {\bibfnamefont {Wei}\ \bibnamefont {Cui}}, \bibinfo
  {author} {\bibfnamefont {J.~Q.}\ \bibnamefont {You}}, \ and\ \bibinfo
  {author} {\bibfnamefont {Franco}\ \bibnamefont {Nori}},\ }\bibfield  {title}
  {\enquote {\bibinfo {title} {Quantum memory using a hybrid circuit with flux
  qubits and nitrogen-vacancy centers},}\ }\href@noop {} {\bibfield  {journal}
  {\bibinfo  {journal} {Phys.\ Rev. A}\ }\textbf {\bibinfo {volume} {88}},\
  \bibinfo {pages} {012329} (\bibinfo {year} {2013})}\BibitemShut {NoStop}%
\bibitem [{\citenamefont {Balasubramanian}\ \emph {et~al.}(2009)\citenamefont
  {Balasubramanian}, \citenamefont {Neumann}, \citenamefont {Twitchen},
  \citenamefont {Markham}, \citenamefont {Kolesov}, \citenamefont {Mizuochi},
  \citenamefont {Isoya}, \citenamefont {Achard}, \citenamefont {Beck},
  \citenamefont {Tissler}, \citenamefont {Jacques}, \citenamefont {Hemmer},
  \citenamefont {Jelezko},\ and\ \citenamefont {Wrachtrup}}]{Naure-Mat}%
  \BibitemOpen
  \bibfield  {author} {\bibinfo {author} {\bibfnamefont {G.}~\bibnamefont
  {Balasubramanian}}, \bibinfo {author} {\bibfnamefont {P.}~\bibnamefont
  {Neumann}}, \bibinfo {author} {\bibfnamefont {D.}~\bibnamefont {Twitchen}},
  \bibinfo {author} {\bibfnamefont {M.}~\bibnamefont {Markham}}, \bibinfo
  {author} {\bibfnamefont {R.}~\bibnamefont {Kolesov}}, \bibinfo {author}
  {\bibfnamefont {N.}~\bibnamefont {Mizuochi}}, \bibinfo {author}
  {\bibfnamefont {J.}~\bibnamefont {Isoya}}, \bibinfo {author} {\bibfnamefont
  {J.}~\bibnamefont {Achard}}, \bibinfo {author} {\bibfnamefont
  {J.}~\bibnamefont {Beck}}, \bibinfo {author} {\bibfnamefont {J.}~\bibnamefont
  {Tissler}}, \bibinfo {author} {\bibfnamefont {V.}~\bibnamefont {Jacques}},
  \bibinfo {author} {\bibfnamefont {P.~R.}\ \bibnamefont {Hemmer}}, \bibinfo
  {author} {\bibfnamefont {F.}~\bibnamefont {Jelezko}}, \ and\ \bibinfo
  {author} {\bibfnamefont {J.}~\bibnamefont {Wrachtrup}},\ }\bibfield  {title}
  {\enquote {\bibinfo {title} {Ultralong spin coherence time in isotopically
  engineered diamond},}\ }\href@noop {} {\bibfield  {journal} {\bibinfo
  {journal} {Nature Mater.}\ }\textbf {\bibinfo {volume} {8}},\ \bibinfo
  {pages} {383} (\bibinfo {year} {2009})}\BibitemShut {NoStop}%
\bibitem [{\citenamefont {Teufel}\ \emph {et~al.}(2011)\citenamefont {Teufel},
  \citenamefont {Donner}, \citenamefont {Li}, \citenamefont {Harlow},
  \citenamefont {Allman}, \citenamefont {Cicak}, \citenamefont {Sirois},
  \citenamefont {Whittaker}, \citenamefont {Lehnert},\ and\ \citenamefont
  {Simmonds}}]{nature-475-359}%
  \BibitemOpen
  \bibfield  {author} {\bibinfo {author} {\bibfnamefont {J.~D.}\ \bibnamefont
  {Teufel}}, \bibinfo {author} {\bibfnamefont {T.}~\bibnamefont {Donner}},
  \bibinfo {author} {\bibfnamefont {Dale}\ \bibnamefont {Li}}, \bibinfo
  {author} {\bibfnamefont {J.~W.}\ \bibnamefont {Harlow}}, \bibinfo {author}
  {\bibfnamefont {M.~S.}\ \bibnamefont {Allman}}, \bibinfo {author}
  {\bibfnamefont {K.}~\bibnamefont {Cicak}}, \bibinfo {author} {\bibfnamefont
  {A.~J.}\ \bibnamefont {Sirois}}, \bibinfo {author} {\bibfnamefont {J.~D.}\
  \bibnamefont {Whittaker}}, \bibinfo {author} {\bibfnamefont {K.~W.}\
  \bibnamefont {Lehnert}}, \ and\ \bibinfo {author} {\bibfnamefont {R.~W.}\
  \bibnamefont {Simmonds}},\ }\bibfield  {title} {\enquote {\bibinfo {title}
  {Sideband cooling of micromechanical motion to the quantum ground state},}\
  }\href@noop {} {\bibfield  {journal} {\bibinfo  {journal} {Nature}\ }\textbf
  {\bibinfo {volume} {475}},\ \bibinfo {pages} {359} (\bibinfo {year}
  {2011})}\BibitemShut {NoStop}%
\bibitem [{\citenamefont {Chan}\ \emph {et~al.}(2011)\citenamefont {Chan},
  \citenamefont {Alegre}, \citenamefont {Safavi-Naeini}, \citenamefont {Hill},
  \citenamefont {Krause}, \citenamefont {Groblacher}, \citenamefont
  {Aspelmeyer},\ and\ \citenamefont {Painter}}]{nature-478-89}%
  \BibitemOpen
  \bibfield  {author} {\bibinfo {author} {\bibfnamefont {Jasper}\ \bibnamefont
  {Chan}}, \bibinfo {author} {\bibfnamefont {T.~P.~Mayer}\ \bibnamefont
  {Alegre}}, \bibinfo {author} {\bibfnamefont {Amir~H.}\ \bibnamefont
  {Safavi-Naeini}}, \bibinfo {author} {\bibfnamefont {Jeff~T.}\ \bibnamefont
  {Hill}}, \bibinfo {author} {\bibfnamefont {Alex}\ \bibnamefont {Krause}},
  \bibinfo {author} {\bibfnamefont {Simon}\ \bibnamefont {Gr\"{o}blacher}},
  \bibinfo {author} {\bibfnamefont {Markus}\ \bibnamefont {Aspelmeyer}}, \ and\
  \bibinfo {author} {\bibfnamefont {Oskar}\ \bibnamefont {Painter}},\
  }\bibfield  {title} {\enquote {\bibinfo {title} {Laser cooling of a
  nanomechanical oscillator into its quantum ground state},}\ }\href@noop {}
  {\bibfield  {journal} {\bibinfo  {journal} {Nature}\ }\textbf {\bibinfo
  {volume} {478}},\ \bibinfo {pages} {89} (\bibinfo {year} {2011})}\BibitemShut
  {NoStop}%
\bibitem [{\citenamefont {Johansson}\ \emph {et~al.}(2013)\citenamefont
  {Johansson}, \citenamefont {Nation},\ and\ \citenamefont {Nori}}]{CPC}%
  \BibitemOpen
  \bibfield  {author} {\bibinfo {author} {\bibfnamefont {J.}~\bibnamefont
  {Johansson}}, \bibinfo {author} {\bibfnamefont {P.}~\bibnamefont {Nation}}, \
  and\ \bibinfo {author} {\bibfnamefont {F.}~\bibnamefont {Nori}},\ }\bibfield
  {title} {\enquote {\bibinfo {title} {Qutip 2: A python framework for the
  dynamics of open quantum systems},}\ }\href@noop {} {\bibfield  {journal}
  {\bibinfo  {journal} {Comput. Phys. Commun.}\ }\textbf {\bibinfo {volume}
  {184}},\ \bibinfo {pages} {1234} (\bibinfo {year} {2013})}\BibitemShut
  {NoStop}%
\end{thebibliography}

\begin{thebibliography}{12}%
\makeatletter
\providecommand \@ifxundefined [1]{%
 \@ifx{#1\undefined}
}%
\providecommand \@ifnum [1]{%
 \ifnum #1\expandafter \@firstoftwo
 \else \expandafter \@secondoftwo
 \fi
}%
\providecommand \@ifx [1]{%
 \ifx #1\expandafter \@firstoftwo
 \else \expandafter \@secondoftwo
 \fi
}%
\providecommand \natexlab [1]{#1}%
\providecommand \enquote  [1]{``#1''}%
\providecommand \bibnamefont  [1]{#1}%
\providecommand \bibfnamefont [1]{#1}%
\providecommand \citenamefont [1]{#1}%
\providecommand \href@noop [0]{\@secondoftwo}%
\providecommand \href [0]{\begingroup \@sanitize@url \@href}%
\providecommand \@href[1]{\@@startlink{#1}\@@href}%
\providecommand \@@href[1]{\endgroup#1\@@endlink}%
\providecommand \@sanitize@url [0]{\catcode `\\12\catcode `\$12\catcode
  `\&12\catcode `\#12\catcode `\^12\catcode `\_12\catcode `\%12\relax}%
\providecommand \@@startlink[1]{}%
\providecommand \@@endlink[0]{}%
\providecommand \url  [0]{\begingroup\@sanitize@url \@url }%
\providecommand \@url [1]{\endgroup\@href {#1}{\urlprefix }}%
\providecommand \urlprefix  [0]{URL }%
\providecommand \Eprint [0]{\href }%
\providecommand \doibase [0]{http://dx.doi.org/}%
\providecommand \selectlanguage [0]{\@gobble}%
\providecommand \bibinfo  [0]{\@secondoftwo}%
\providecommand \bibfield  [0]{\@secondoftwo}%
\providecommand \translation [1]{[#1]}%
\providecommand \BibitemOpen [0]{}%
\providecommand \bibitemStop [0]{}%
\providecommand \bibitemNoStop [0]{.\EOS\space}%
\providecommand \EOS [0]{\spacefactor3000\relax}%
\providecommand \BibitemShut  [1]{\csname bibitem#1\endcsname}%
\let\auto@bib@innerbib\@empty
\bibitem{book-1} L. D. Landau and E. M. Lifshitz, \emph{Theory of Elasticity} (Butterworth-Heinemann, Oxford, 1986).
\bibitem{Phys.Rep-1} Menno Poot and Herre S. J. van der Zant, Phys. Rep. \textbf{511}, 273 (2012).
\bibitem{prb-67-235414} S. Sapmaz, Ya. M. Blanter, L. Gurevich, and H. S. J. van der Zant, Phys. Rev. B \textbf{67}, 235414 (2003).
\bibitem{nl-14-1} Mehmet Aykol, Bingya Hou, Rohan Dhall, Shun-Wen Chang, William Branham, Jing Qiu, and Stephen B. Cronin, Nano Lett. \textbf{14}, 2426 (2014).
\bibitem{prb-rabl} P. Rabl, P. Cappellaro, M. V. Gurudev Dutt, L. Jiang,  J. R. Maze, and M. D. Lukin, Phys. Rev. B \textbf{79}, 041302 (R) (2009).
\bibitem{NJP-14 115004} S. J. M. Habraken, K. Stannigel, M. D. Lukin, P. Zoller, and P. Rabl, New J. Phys. \textbf{14} 115004 (2012).
\bibitem{prl-95-076803} S. Roche,  Jie Jiang,  F. Triozon,  and R. Saito, Phys. Rev. Lett. \textbf{95}, 076803 (2005).
\bibitem{JAP-98-124301} Q. Wang, J. Appl. Phys. \textbf{98}, 124301 (2005).
\bibitem{physica-E-40-2791} H. Heireche, A. Tounsi, A. Benzair,  M. Maachou, E.A. Adda Bedia, Physica E \textbf{40}, 2791(2008).
\bibitem{Na-Nano} J. Moser, A. Eichler, J. G\"{u}ttinger, M. I. Dykman, and A. Bachtold,  Nat. Nanotech. \textbf{9}, 1007 (2014).

\end{thebibliography}

 %

\onecolumngrid

\appendix

\clearpage

\section*{Supplemental Material:  }


\setcounter{equation}{0}
\setcounter{figure}{0}
\setcounter{table}{0}
\setcounter{page}{1}
\makeatletter
\renewcommand{\theequation}{S\arabic{equation}}
\renewcommand{\thefigure}{S\arabic{figure}}
\renewcommand{\bibnumfmt}[1]{[S#1]}
\renewcommand{\citenumfont}[1]{S#1}

\subsection{Fundamental vibration mode of the carbon nanotube}

The motion of a suspended nanotube can be described by the Euler-Bernoulli theory \cite{book-1,Phys.Rep-1}. The Euler-Bernoulli equation for the static and dynamic displacement of a thin beam subjected to an external driving force reads
\begin{eqnarray}
  \rho A\frac{\partial^2}{\partial t^2 }\phi(x,t)+E\mathcal {I}\frac{\partial^4}{\partial x^4 }\phi(x,t)-T\frac{\partial^2}{\partial x^2 }\phi(x,t)&=&F_\text{ext}(x,t),
\end{eqnarray}
where $\phi(x,t)$ is the lateral displacement in the $y$ direction, $\rho$ the mass density, $A$  the beam cross section, $E$ the Young modulus, $\mathcal {I}$  the moment of inertia, $T$  the tension in the tube, and $F_\text{ext}(x,t)$
a unit length force that accounts for the effect of the gate electrodes. The
displacement amplitude increases remarkably when the tube
is driven at an eigenfrequency of the system. These eigenfrequencies
correspond to the bending modes of the nanotube,
which we label by mode number $n$ starting from zero for
the fundamental mode. We here consider the eigenfrequencies of a perfect clamping nanotube resonator, in which case the tension $T$ goes to zero at zero gate voltage $F_\text{ext}(x,t)=0$ \cite{prb-67-235414,nl-14-1}. In this case the Euler-Bernoulli equation reads
\begin{eqnarray}
  \rho A\frac{\partial^2}{\partial t^2 }\phi(x,t)+E\mathcal {I}\frac{\partial^4}{\partial x^4 }\phi(x,t)&=&0.
\end{eqnarray}
The eigenmodes $\phi_{n}$ and eigenfrequencies $\omega_{n}$ satisfy:
\begin{eqnarray}
\omega_n^2\rho A\phi_{n}&=&E\mathcal {I}\frac{\partial^4}{\partial x^4 }\phi_n
\end{eqnarray}
The solutions to this equation are
\begin{eqnarray}
 \phi_n(x) &=& C_1(\cos k_nx-\cosh k_nx)+C_2(\sin k_nx-\sinh k_nx).
\end{eqnarray}
For a doubly clamped beam,  the boundary conditions are $\phi_n(0)=\phi_n(L)=0,\phi_n'(0)=\phi_n'(L)=0$, and for a cantilever
the boundary conditions are $\phi_n(0)=\phi_n'(0)=0,\phi_n''(L)=\phi_n'''(L)=0$.
For the former case, the frequency equation is given by
\begin{equation}
  \cos k_nL\cosh k_nL=1.
\end{equation}
For the latter, the frequency equation is given by
\begin{equation}
  \cos k_nL\cosh k_nL=-1.
\end{equation}
The first five nontrivial consecutive roots of these equations are given
below
\begin{ruledtabular}
\begin{tabular}{ccc}
Mode & Cantilever    & Beam \\
n& $k_nL$&$k_nL$\\
\colrule
0&  $1.875$&$ 4.730$\\
1&  $4.694$&$ 7.853$\\
2&  $7.855$&$10.996 $\\
3&  $10.996$&$14.137$\\
4&  $14.137$&$17.279$
\end{tabular}
\end{ruledtabular}
The corresponding eigenfrequencies are
\begin{equation}
\omega_\text{n}=k_n^2\sqrt{\frac{E\mathcal {I}}{\rho A}}.
\end{equation}
Therefore, the fundamental vibrational mode of a nanotube has the vibration frequency  $\omega_\text{nt}\sim\frac{1}{L^2}\sqrt{\frac{E\mathcal {I}}{\rho A}}$ \cite{prb-67-235414,nl-14-1}.
In table \ref{tab:table1} we present the relevant parameters for the carbon nanotube resonator without gate voltages.
\begin{table}
\caption{\label{tab:table1}%
Parameters for the nanotube considered in this work.
}
\begin{ruledtabular}
\begin{tabular}{llc}
Term &
Value  &
Units \\
\colrule
Length $L$ & 2 & $\mu$m\\
Radius $r$& 1.5 & nm\\
Wall thickness $t$ & 0.335 & nm\\
Mass density $\rho$ & 1350  &  kg/$\text{m}^3$\\
Effective mass  $m$ & $7\times10^{-21}$ &  kg\\
Young
modulus $E$& 1 & TPa\\
Fundamental frequency $\omega_0$ & $2\pi\times$2  & MHz\\
Current carrying capacity $C$   & $\geq10$ & $\mu$A/$\text{nm}^2$\\
Current $I$ & 60 & $\mu$A\\
Quality factor $Q$  &$10^5$ &   /
\end{tabular}
\end{ruledtabular}
\end{table}

\subsection{Derivation of the spin-vibration interaction}

The interaction
of a  single NV center located at $\vec{r}$
with  the  total magnetic field (external driving and from the nanotube) is
\begin{eqnarray}
\hat{H}_\text{NV}&=&\hbar D \hat{S}_z^2+\mu_Bg_s\ B_z\hat{S}_z+\mu_Bg_s (\vec{B}_{\text{nt}}(\vec{r})+ \vec{B}_\text{dr})\cdot\hat{\vec{S}}.
\end{eqnarray}
Expanding the magnetic field $\vec{B}_{\text{nt}}(\vec{r})$ up to first order in $\hat{u}_y$, we have
\begin{eqnarray}
\hat{H}_\text{NV}&=&\hbar D \hat{S}_z^2+\mu_Bg_s[\ B_z+B_\text{nt}(d)]\hat{S}_z+\mu_Bg_s   \vec{B}_\text{dr}\cdot\hat{\vec{S}}+\mu_Bg_s \hat{S}_z\partial_yB_\text{nt}\hat{u}_y.
\end{eqnarray}
In the basis defined by the eigenstates of $\hat{S}_z$, i.e., $\{\vert m_s\rangle,m_s=0,\pm1\}$, with $\hat{S}_z\vert m_s\rangle=m_s\vert m_s\rangle$, we get
\begin{eqnarray}
\hat{H}_\text{NV}&=&\sum_{m_s}\{ \langle m_s\vert[\hbar D \hat{S}_z^2+\mu_Bg_s[\ B_z+B_\text{nt}(d)]\hat{S}_z]\vert m_s\rangle\}\vert m_s\rangle \langle m_s\vert\nonumber \\
&&+\sum_{m_s,m'_s}\{ \langle m_s\vert\mu_Bg_s\vec{B}_\text{dr}\cdot\hat{\vec{S}}\vert m'_s\rangle\}\vert m_s\rangle \langle m'_s\vert
+\sum_{m_s}\{ \langle m_s\vert\mu_Bg_s \hat{S}_z\partial_yB_\text{nt}\hat{u}_y\vert m_s\rangle\}\vert m_s\rangle \langle m_s\vert.
\end{eqnarray}
Taking $\vec{B}_\text{dr}=B_\text{0}\cos\omega_0 t\vec{e}_x=B_0/2(e^{i\omega_0 t}+e^{-i\omega_0 t})\vec{e}_x$, we have
\begin{eqnarray}
\hat{H}_\text{NV}&=&\sum_{m_s}\{\hbar D m_s^2+\mu_Bg_s[ B_z+B_\text{nt}(d)]m_s\}\vert m_s\rangle \langle m_s\vert\nonumber \\
&&+\sum_{m_s,m'_s}\frac{1}{2}\mu_Bg_sB_0(e^{i\omega_0 t}+e^{-i\omega_0 t}) \langle m_s\vert\hat{S}_x\vert m'_s\rangle\vert m_s\rangle \langle m'_s\vert\nonumber \\
&&+\sum_{m_s}\mu_Bg_s(\hbar/2m\omega_\text{nt})^{1/2}\partial_yB_\text{nt}m_s\vert m_s\rangle \langle m_s\vert(\hat{a}^\dag+\hat{a}).
\end{eqnarray}
In the rotating-frame at the driving frequency $\omega_0$ and under the rotating-wave approximation, we can obtain
\begin{eqnarray}
\hat{H}_\text{NV}&=&(\hbar D +\mu_Bg_sB_z+B_\text{nt}-\hbar\omega_0)\vert +1\rangle \langle +1\vert+(\hbar D -\mu_Bg_sB_z-B_\text{nt}-\hbar\omega_0)\vert -1\rangle \langle -1\vert\nonumber \\
&&+\frac{\sqrt{2}}{4}\mu_Bg_sB_0 (\vert 0\rangle \langle +1\vert+\vert +1\rangle \langle 0\vert)+\frac{\sqrt{2}}{4}\mu_Bg_sB_0 (\vert 0\rangle \langle -1\vert+\vert -1\rangle \langle 0\vert)\nonumber \\
&&+\mu_Bg_s(\hbar/2m\omega_\text{nt})^{1/2}\partial_yB_\text{nt}(\vert +1\rangle \langle +1\vert-\vert -1\rangle \langle -1\vert)(\hat{a}^\dag+\hat{a}).
\end{eqnarray}
Including the free Hamiltonian of the vibration mode, we have
\begin{eqnarray}\label{SH1}
\hat{H}_\text{NV}&=&\hbar \omega_\text{nt} \hat{a}^\dag\hat{a}+\hbar \Delta_+ \vert+1\rangle\langle+1\vert+ \hbar\Delta_-\vert-1\rangle\langle-1\vert\nonumber\\
&&+\hbar \Omega[\vert-1\rangle\langle0\vert+\vert0\rangle\langle-1\vert] +\hbar \Omega[\vert+1\rangle\langle0\vert+\vert0\rangle\langle+1\vert]\nonumber\\
&&+\hbar g( \vert+1\rangle\langle+1\vert-\vert-1\rangle\langle-1\vert)(\hat{a}^\dag+\hat{a}).
\end{eqnarray}
with $\hbar\Delta_\pm=\hbar D\pm\mu_Bg_s(B_z+B_\text{nt})-\hbar \omega_0$, $\hbar \Omega=\frac{\sqrt{2}}{4}\mu_Bg_sB_0$, and $\hbar g=\mu_Bg_s(\hbar/2m\omega_\text{nt})^{1/2}\partial_yB_\text{nt}$.

In the following we assume symmetric detunings $\Delta_+=\Delta_-=\Delta$ for simplicity. We can define the bright and dark states for the NV spin states
\begin{eqnarray}
\vert \mathcal {B}\rangle&=&\frac{1}{\sqrt{2}}(\vert +1\rangle+\vert-1\rangle)\nonumber\\
\vert \mathcal {D}\rangle&=&\frac{1}{\sqrt{2}}(\vert +1\rangle-\vert-1\rangle).
\end{eqnarray}
Then we find that the microwave field couples the state $\vert0\rangle$ to the bright state $\vert \mathcal {B}\rangle$, while the dark  state $\vert \mathcal {D}\rangle$ remains decoupled. In the dressed state basis $\{ \vert \mathcal{G}\rangle=\cos\theta\vert0\rangle-\sin\theta\vert \mathcal {B}\rangle, \vert \mathcal{E}\rangle=\cos\theta\vert\mathcal {B}\rangle+\sin\theta\vert 0\rangle\}$,with $\tan2\theta=2\sqrt{2}\Omega/\Delta$, Hamiltonian (\ref{SH1}) can be rewritten as \cite{prb-rabl}
\begin{eqnarray}\label{SH2}
\hat{H}_\text{NV}&=&\hbar \omega_\text{nt} \hat{a}^\dag\hat{a}+\hbar \omega_{eg} \vert  \mathcal{E}\rangle\langle \mathcal{E}\vert+ \hbar\omega_{dg}\vert \mathcal{D}\rangle\langle \mathcal{D}\vert\nonumber\\
&&+\hbar (g_1 \vert \mathcal{G}\rangle\langle \mathcal{D}\vert+g_2\vert \mathcal{D}\rangle\langle \mathcal{E}\vert+\text{H.c.})(\hat{a}^\dag+\hat{a}),
\end{eqnarray}
where $\omega_{eg}=\sqrt{\Delta^2+8\Omega^2}$, $\omega_{dg}=\frac{\Delta+\sqrt{\Delta^2+8\Omega^2}}{2}$, $g_1=-g\sin\theta$, and $g_2=g\cos\theta$.
Under the condition $\Delta\gg \Omega$, one has $\sin\theta\simeq 0$, $\cos\theta\simeq 1$, $\omega_{eg}\simeq\Delta+\frac{4\Omega^2}{\Delta}$, $\omega_{dg}\simeq\Delta+\frac{2\Omega^2}{\Delta}$, and $\vert \mathcal{E}\rangle\simeq \vert \mathcal{B}\rangle$, which
leads to
\begin{eqnarray}
\hat{\mathcal{H}}_q&=&\hbar \omega_\text{nt} \hat{a}^\dag\hat{a}+\frac{1}{2}\hbar \Lambda \hat{\sigma}_z+\hbar g(  \hat{\sigma}_++\hat{\sigma}_-)(\hat{a}^\dag+\hat{a}).
\end{eqnarray}

\subsection{The two-qubit operations}

\subsubsection{\textbf{Strong spin-spin interactions mediated by phonons}}
We consider two  NV centers, separated by a distance $l\sim 1\mu m$,  coupled to the same vibration mode of the nanotube, with  Hamiltonian
\begin{eqnarray}
\hat{\mathcal{H}}_{2q}&=&\hbar \omega_\text{nt} \hat{a}^\dag\hat{a}+\sum_{i=1,2}\frac{1}{2}\hbar \Lambda_i \hat{\sigma}_z^{i}
+\sum_{i=1,2}\hbar g_i(  \hat{\sigma}_+^i+\hat{\sigma}_-^i)(\hat{a}^\dag+\hat{a}).
\end{eqnarray}
For simplicity, we assume $\Lambda_1\simeq\Lambda_2=\Lambda$ and $g_1\simeq g_2=g$, and
 consider the dispersive regime $|\Lambda- \omega_\text{nt}|\gg g$, when two NV centers are far detuned from the resonator
but in resonance with each other.  After the use of  a Schrieffer-Wolff transformation, this will lead to an effective nonlocal spin-spin interaction via the exchange of virtual phonons,
\begin{eqnarray}
\hat{\mathcal{H}}_\text{s-s}&=& \hbar \lambda_\text{eff} (\hat{\sigma}_+^1\hat{\sigma}_-^2+\hat{\sigma}_-^1\hat{\sigma}_+^2),
\end{eqnarray}
with the coupling strength $\lambda_\text{eff}=g^2/|\Lambda- \omega_\text{nt}|$. This interaction can extend over
distances on the order of the nanotube's length, which allows us to coherently control the interactions between distant NV spins.

We now proceed to discuss the coherence length $l_c$ of the phonon mediated NV spin coupling, which is  essential to evaluate the application potential of this device. In particular, it is a very important issue when propagating phonons are considered rather than the confined one used in this work.
At ambient temperature, phonon scattering inside solid state materials results in harmful decoherence processes. An important figure of merit that is used to quantitatively characterize all the phonon dissipation mechanisms, including the decay of vibrations into the support as well as intrinsic damping mechanism within the nanotube, e.g. due to scattering from surface defects,  is the mechanical quality factor  $Q$, defined as   the ratio of the resonant frequency over the linewidth.
It is  interesting to estimate a sort of coherence length for the nanoresonator mode. To do so one can think of the fundamental bending mode as a traveling wave, which is reflected at the ends of the nanotube. Thus, the coherence length
can be estimated as the effective mean phonon free path $l_c\simeq v \tau$ \cite{NJP-14 115004,prl-95-076803}, where $v$ is the effective speed of sound, and $\tau$ is the relaxation time (mechanical damping rate $\gamma_m=\tau^{-1}$).
The effective speed of sound $v$ can be found from the  relation as $\omega_\text{nt}=v k_0$ \cite{JAP-98-124301,physica-E-40-2791}, while the relaxation time is related to the quality factor of the mechanical mode $\frac{1}{\tau}=\omega _\text{nt}/Q$. Thus, we can estimate the coherence length of the phonon mediated NV spin coupling as $l_c\sim Q/k_0\sim QL$. It can extend over distances on the order of several  centimeters, much larger than the distance between two NV spins. Therefore, for a high-Q mechanical resonator, phonons can coherently propagate inside it back and forth for quite a long distance before they finally dissipate. This phenomenon has a direct analogy to photons bound in a high-Q
micro-cavity. Thus, in this scheme, we can safely ignore the harmful effect of phonon scattering inside solid state materials, provided that the mechanical resonator
possesses a very high quality factor at low temperature.  We need to emphasize that the recent fabrication of carbon nanotube resonators can possess a quality
factor exceeding $10^5$ \cite{Na-Nano}, which ensures that phonon losses do not severely limit our scheme. The minimum requirement of this work is that the length scale of the phonon mediated NV spin coupling is on the order of the tube's length, which would allow us to coherently control the NV spin interactions, and  facilitate potential applications of this hybrid device.

\subsubsection{\textbf{Dynamics of the two coupled NV spins} }
To implement this protocol, we need a microwave to drive  the transition between the qubit state $\vert 0\rangle_j$ and the bright state $\vert\mathcal {B}\rangle_j$ in each qubit with Rabi frequency $\Omega_j$ and frequency detuning $\delta_j$. The dynamics of the entire system is described by
 \begin{eqnarray}
\hat{\mathcal{H}}=\sum_j\hbar \delta_j \vert \mathcal{B}\rangle_{jj}\langle \mathcal{B}\vert+\sum_j[{\hbar \Omega_j\vert \mathcal{B}\rangle_{jj}\langle 0\vert +\text{H.c.}}]+\hat{\mathcal{H}}_\text{s-s}.
\end{eqnarray}
 The spin-spin interaction can be diagonalized with the states $\vert \pm\rangle_q=1/\sqrt{2}[\vert \mathcal{B}\rangle_1\vert\mathcal{D}\rangle_2\pm\vert\mathcal{D}\rangle_1\vert \mathcal{B}\rangle_2]$, leading to
\begin{eqnarray}
 \hat{\mathcal{H}}_\text{s-s}=\hbar \lambda_\text{eff} \vert+\rangle_{qq}\langle+\vert-\hbar \lambda_\text{eff} \vert-\rangle_{qq}\langle-\vert.
\end{eqnarray}

To implement a SWAP gate and quantum
information transfer between two qubits, we encode quantum information in the two spin states as $\vert 0\rangle_q=\vert0\rangle$ and $\vert 1\rangle_q=\vert \mathcal{D}\rangle$. The entire system is described by
\begin{eqnarray}\label{SH3}
\hat{\mathcal{H}}=\sum_j\hbar \delta_j \vert \mathcal{B}\rangle_{jj}\langle \mathcal{B}\vert+\sum_j[{\hbar \Omega_j\vert \mathcal{B}\rangle_{jj}\langle 0\vert +\text{H.c.}}]+\hbar \lambda_\text{eff} \vert+\rangle_{qq}\langle+\vert-\hbar \lambda_\text{eff} \vert-\rangle_{qq}\langle-\vert.
\end{eqnarray}
To gain more insight into the dynamics of the coupled system, we write the Hamiltonian
Eq. (\ref{SH3}) in the space $S$ spanned by the state vectors $\{\vert 0,1\rangle_q,\vert+\rangle_q,\vert-\rangle_q,\vert
1,0\rangle_q,\vert0,0\rangle_q,\vert1,1\rangle_q,\vert \mathcal{B},\mathcal{B}\rangle_q,\vert 0,\mathcal{B}\rangle_q,\vert \mathcal{B},0\rangle_q \}$,
\begin{eqnarray}
\hat{\mathcal{H}}=\hbar\left[
  \begin{array}{ccccccccc}
    0 & \bar{\Omega}_1 &\bar{\Omega}_1& 0 & 0 & 0 & 0 & 0 & 0 \\
    \bar{\Omega}_1& \delta_+ & 0 & \bar{\Omega}_2 & 0 & 0 & 0 & 0 & 0 \\
   \bar{\Omega}_1 & 0 & -\delta_- & -\bar{\Omega}_2 & 0 & 0 & 0 & 0 & 0 \\
    0 & \bar{\Omega}_2 & -\bar{\Omega}_2 & 0 & 0 & 0 &0 & 0 & 0 \\
    0 & 0 & 0 & 0 & 0 & 0 & 0 & \Omega_2 & \Omega_1 \\
    0 & 0 & 0 & 0 & 0 & 0 & 0 & 0 & 0 \\
    0 & 0 & 0 & 0 & 0 & 0 & \delta_1+\delta_2 & \Omega_1 & \Omega_2 \\
    0 & 0 & 0 & 0 & \Omega_2 & 0 & \Omega_1 & \delta_2 & 0 \\
    0 & 0 & 0 & 0 & \Omega_1 & 0 & \Omega_2 & 0 & \delta_1 \\
  \end{array}
\right]
\end{eqnarray}
From the matrix form for the Hamiltonian
Eq. (\ref{SH3}), we see that the space $S$ can be decomposed into two independent subspaces $S_1=\{\vert 0,1\rangle_q,\vert+\rangle_q,\vert-\rangle_q,\vert
1,0\rangle_q\}$ and $S_2=\{\vert0,0\rangle_q,\vert1,1\rangle_q,\vert \mathcal{B},\mathcal{B}\rangle_q,\vert 0,\mathcal{B}\rangle_q,\vert \mathcal{B},0\rangle_q\}$, i.e., $S=S_1\oplus S_2$.
Thus we can find that if the two qubits are initially prepared in the state $\vert 0\rangle_q^1\vert1\rangle_q^2$ or $\vert 1\rangle_q^1\vert0\rangle_q^2$, then the dynamics of the system will be confined in the subspace $S_1$ governed by the Hamiltonian
\begin{eqnarray}
  \hat{\mathcal{H}} &=& \hbar\delta_+\vert+\rangle_{qq}\langle+\vert-\hbar\delta_-\vert-\rangle_{qq}\langle-\vert +\hbar\bar{\Omega}_1\vert+\rangle_{qq}\langle0,1\vert \nonumber \\
  &&+\hbar\bar{\Omega}_1\vert-\rangle_{qq}\langle0,1\vert+\hbar\bar{\Omega}_2\vert+\rangle_{qq}\langle1,0\vert
 -\hbar\bar{\Omega}_2\vert-\rangle_{qq}\langle1,0\vert+\text{H.c.}
\end{eqnarray}


%

\end{document}